\documentclass[sigconf]{acmart}

\usepackage{amsmath, amsfonts}
\usepackage{graphicx} 
\usepackage{caption, subcaption} 
\usepackage{algorithm} 
\usepackage{algpseudocode}
\usepackage{float} 
\usepackage{textcomp} 
\usepackage{multirow} 
\usepackage{tabularray} 
\usepackage{caption}
\usepackage{hyperref}
\usepackage{enumitem}
\copyrightyear{2025}
\acmYear{2025}
\setcopyright{cc}
\setcctype{by}
\acmConference[UMAP '25]{33rd ACM Conference on User Modeling, Adaptation and Personalization}{June 16--19, 2025}{New York City, NY, USA}
\acmBooktitle{33rd ACM Conference on User Modeling, Adaptation and Personalization (UMAP '25), June 16--19, 2025, New York City, NY, USA}
\acmDOI{10.1145/3699682.3728340}
\acmISBN{979-8-4007-1313-2/2025/06}

\begin{document}

\newcommand{\namei}{{\em Memento}}
\newcommand{\name}{{\em Memento }}
\newcommand{\names}{{\em Memento}}
\newcommand{\namef}{{\em Memento's\ }}
\newcommand{\namet}{Memento}
\newcommand{\nameb}{Memento }

\author{Indrajeet Ghosh$^1$, Kasthuri Jayarajah$^2$, Nicholas Waytowich{$^3$}, Nirmalya Roy$^1$ \\
$^1$Information Systems, UMBC, $^2$Computer Science, NJIT, {$^3$}DEVCOM, Army Research Lab, USA}
 \email{indrajeetghosh@umbc.edu,  kasthuri.jayarajah@njit.edu,  nicholas.r.waytowich.civ@army.mil,  nroy@umbc.edu}

\title{\textit{Memento:} Augmenting Personalized Memory via Practical Multimodal Wearable Sensing in Visual Search and Wayfinding Navigation}

\renewcommand{\shortauthors}{Ghosh et al.}

\begin{abstract}
\textit{Working memory} involves the temporary retention of information over short periods. It is a critical cognitive function that enables humans to perform various online processing tasks, such as dialing a phone number, recalling misplaced items' locations, or navigating through a store. However, inherent limitations in an individual’s capacity to retain information often result in forgetting important details during such tasks. Although previous research has successfully utilized wearable and assistive technologies to enhance long-term memory functions (e.g., episodic memory), their application to supporting short-term recall in daily activities remains underexplored. To address this gap, we present \namei, a framework that uses multimodal wearable sensor data to detect significant changes in cognitive state and provide intelligent in situ cues to enhance recall. Through two user studies involving \textbf{15} and \textbf{25} participants in visual search navigation tasks, we demonstrate that participants receiving visual cues from \namei~achieved significantly better route recall, improving approximately \textbf{ 20-23\%} compared to free recall. Furthermore, \namei~reduced cognitive load and review time by \textbf{46\%} while also achieving a substantial reduction in computation time (\textit{3.86 secs} vs. \textit{15.35 secs}), offering an average \textbf{75\%} effective compared to computer vision-based cues selection approaches.

\end{abstract}



\maketitle

\section{Introduction}\label{sec:intro}

\textbf{Working Memory (WM)} refers to the temporary retention of information for short periods, enabling humans to process information \emph{online}~\cite{ma2014changing, barrouillet2012time, barrouillet2021time}. It supports tasks such as dialing a phone number, recalling where keys were placed, or deciding which turn to take in a mall. Unlike long-term or episodic memory, which involves storing and retrieving information over extended periods (e.g., years) and is shaped by life experiences, WM focuses on maintaining and manipulating information momentarily. Individual differences in WM capacity—how much one can hold—affect performance and contribute to \emph{forgetting}~\cite{lewis2014competition, aslan2011individual}.

The rise of pervasive sensing technologies (e.g., smartphone, wearable cameras, social media logs) has spurred research on augmenting human memory~\cite{alle2017wearable, davies2015security, dingler2021memory, kandappu2021privacyprimer, gonccalves2004describing, blanc2007people}. While studies have explored assistive technologies like lifelogging for episodic memory~\cite{kandappu2021privacyprimer} and marker-based tracking for older adults~\cite{li2019fmt}, they primarily enhance long-term memory (e.g., autobiographical, episodic) via cameras, raising privacy concerns. To the best of our knowledge, no prior work has specifically aimed at enhancing working memory for short-term retention. Through this work, we address the \textbf{\textit{two}} fundamental research questions:

\begin{itemize}[leftmargin=*]
    \item \textbf{RQ1}:~\emph{Can non-invasive multi-modal wearables effectively capture key high-attentional \textbf{event-related potential} (ERP) episodes during working memory tasks?}
    \item \textbf{RQ2}:~\emph{Can capturing and highlighting the high-attentional ERP episodes and intelligently cueing them can improve memory recall of significant personalized moments~("\bf {mementos}")?}
\end{itemize}

\noindent \textbf{Key Challenges:} Although memory augmentation aids effectively enhance recall across various age groups~\cite{le2016impact, hong2017novel}, most techniques rely on lifelogging~\cite{elagroudy2023impact} or continuous ego-centric vision processing, posing significant privacy concerns. While wearable-based physiological sensing—especially EEG—has been explored~\cite{pierce2021wearable, sun2016remembered}, limited spatial and temporal resolution, along with susceptibility to artifacts (e.g., eye blinks, random noise), reduces practical utility~\cite{cao2021unsupervised}. Moreover, artifacts in galvanic skin response~(GSR) and photoplethysmography~(PPG) signals~\cite{cao2021unsupervised, patel2018wearable, rihet2024robot} further compromise these modalities in real-world applications.

\noindent To this end, we propose \names, a multimodal wearable sensing system that enhances human \emph{recall} by integrating functional connectivity-based reconstruction with proactive identification of \emph{moments of significance}. \names~improves signal quality across wearable devices and generates \emph{mementos} of working memory, aiding later retrieval of task-relevant information. \textbf{Our focus is on \textit{visuospatial working memory}}, which plays a crucial role in visual search and wayfinding~\cite{ye2022cognitive, shi2021spatial}. We employ multimodal sensors~(EEG, GSR, and PPG) to detect significant moments without obtrusiveness during the task. Through two user studies (\textbf{15} and \textbf{25} participants), we show that \names~outperform \textbf{six} baseline methods in detecting these moments and improving recall. This work lays the foundation for smart assistive interfaces that augment personalized moment-of-significance for memory augmentation.

In this work, we make the following \emph{Key Research Contributions}:

\begin{itemize}[leftmargin=*]
    \item{\textbf{Physiological Sensing with commercial off-the-shelf (COTS) Devices:} We propose \textit{Memento}, which integrates (a) functional connectivity-based EEG signal reconstruction and (b) \textit{incremental feature-level fusion} with GSR and PPG sensors to enhance the accuracy of COTS EEG devices. This enables effective detection of high-attentional ERP episodes during visuospatial working memory tasks.} 

    \item{\textbf{Multimodal Wearable Sensing for ERPs Extraction:} Through a user study with \textbf{\textit{15}} participants across four environments, we demonstrate that combining EEG reconstruction with multimodal sensing effectively detects high-attentional moments during navigation. \names~successfully extracts up to \textbf{72.6\%} of episodes correlating with subjects' visual attention to object instances—a \textbf{35\%} improvement over raw EEG signals alone.} 

    \item{\textbf{Enhancing Navigation Recall with Wearable-based Visual Cues:} In a separate study with \textbf{\textit{25}} participants, we quantitatively show that \name surpasses \textbf{six} baselines in both navigation retrace recall and cognitive load reduction. Specifically, \name improves recall accuracy by~$\approx{(8-12)\%}$ compared to a computer vision-based approach and~$\approx{(20-23)\%}$ over no memory aid. Moreover, \name yields the lowest PAAS and NASA TLX scores for perceived cognitive load and reduces review time by \textbf{46\%} compared to the CV-based approach. Additionally, the proposed physiological-driven approach achieves a significant reduction in processing complexity, requiring \textbf{3.86 seconds} compared to \textbf{15.35 seconds} per user session, resulting in~$\approx$~\textbf{75\%} reduction.}
\end{itemize}

\section{Related Work}~\label{sec:related work}
This section summarizes the relevant literature in contrast to \textit{Memento} {\it (i)} memory augmentation using pervasive sensing technologies and physiological sensing, memory and cognition in ubiquitous computing. We mainly focus on the aspect of our \name~approach that differentiates from the state-of-the-art~(SOTA) methods.

\noindent \textbf{Memory Augmentation using Pervasive Sensing Technologies.} The rise of pervasive sensing (e.g., smartphones, wearable cameras) has spurred research on memory augmentation~\cite{alle2017wearable, davies2015security, dingler2021memory, silva2013benefits, li2019fmt, kandappu2021privacyprimer}. Davies et al.~\cite{davies2015security} highlight privacy risks of ``total recall'' and potential memory manipulation. Dingler et al.~\cite{dingler2021memory} provide design guidelines for camera-based lifelogging, emphasizing positioning, logging, and summarization. While focused on episodic memory, they underscore cueing’s role in reinforcement, central to our approach. Silva et al.~\cite{silva2013benefits} show that reviewing lifelogged data (e.g., SenseCam) enhances neurophysiological assessments beyond episodic memory, surpassing journaling. Kandappu et al.~\cite{kandappu2021privacyprimer} explore marker-based tracking for memory support in older adults, addressing privacy-preserving lifelogging. Existing studies predominantly target \textit{long-term memory} (e.g., autobiographical, episodic) via camera-based methods, raising privacy concerns. To our knowledge, this work is among the first to (a) enhance short-term memory, specifically working memory, and (b) leverage physiological sensing to capture ERP cues for short-term memory augmentation.

\noindent \textbf{Physiological Sensing, Memory, and Cognition in Ubiquitous Computing.} Wearable-based physiological sensing has been extensively studied in ubiquitous computing, particularly for applications such as emotion recognition in lifelogging~\cite{jiang2019memento}, assessing emotional climate in classrooms~\cite{gashi2018using, di2018unobtrusive}, and promoting well-being~\cite{di2018emotion}. Chwalek et al.~\cite{chwalek2021captivates} have developed sensor integration frameworks to support large-scale user studies. While recent studies have explored cognitive processes~\cite{niforatos2017amplifying, gashi2019using}, cognition-based interventions remain underexplored. Similar to our work in identifying high/low cognitive load through physiological sensing, Prompto~\cite{chan2022prompto} opportunistically delivers prompts when users exhibit lower cognitive load, optimizing learning and memory retention. Bulling and Roggen~\cite{bulling2011recognition} demonstrated the feasibility of recognizing visual memory recall using Electrooculography (EOG) features to distinguish between familiar and unfamiliar images. Our work is the first to extract visual cues from physiological signals to enhance visuospatial working memory recall during navigation tasks.

\section{Designing \names} \label{sec:sysdesign}

\begin{figure}[t]
  \centering
    \includegraphics[scale =0.36]{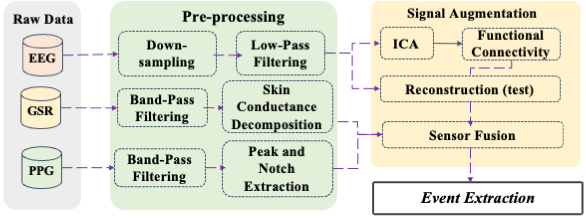}
    \caption{{Overview of the \name Framework} }
    \label{fig:sysoverview}
\end{figure}

\noindent \name~enhances the utility of low-density EEG signals through \textbf{two} key features: (a) functional connectivity-based signal reconstruction and (b) multimodal fusion with auxiliary physiological sensors. This section describes the overall \name~framework, as illustrated in Figure~\ref{fig:sysoverview}. Our multimodal sensing system integrates the Emotiv Insight headset\footnote{\url{https://www.emotiv.com/insight/}} with 5 EEG channels and the Shimmer\footnote{\url{https://shimmersensing.com}} platform for Galvanic Skin Response (GSR) and heart rate (Photoplethysmography, PPG) sensing.

\name~comprises three key modules: (a) \textbf{Data Preprocessing} (Section~\ref{sec:preprocessing}) processes each sensing modality using a suite of signal processing techniques, (b) \textbf{Multimodal Fusion} (Section~\ref{sec:augmentation}) integrates multiple physiological signals, and (c) \textbf{Event Extraction} (Section~\ref{sec:cpd}) detects moments of high attentional focus.

\begin{figure*}
\centering
  \includegraphics[scale=0.22]{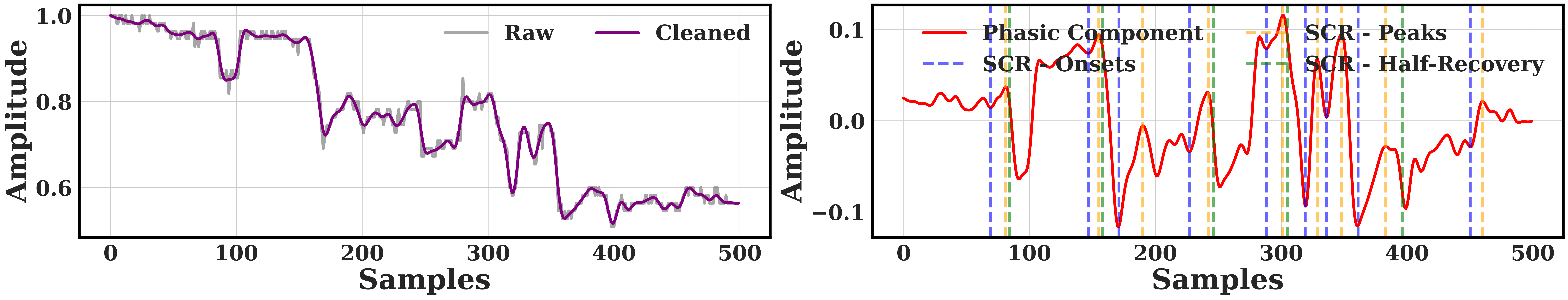}
\caption{An example of preprocessed GSR signal highlighting SCR Peaks, Phasic Component, SCR Onsets, SCR Half Recovery phases and smooth, denoise/clean GSR signals, respectively.}
    \label{fig:gsr_signal}
  \end{figure*}

\subsection{Signal Pre-processing}
\label{sec:preprocessing}

\noindent \textbf{Electroencephalogram (EEG, $\mathcal{E}^{R,C}$).} EEG is a widely used non-invasive neuroimaging modality in cognitive neuroscience~\cite{luo2018effect}. However, COTS EEG devices with sparse electrodes often suffer from degraded sensing fidelity and connectivity~\cite{niso2023wireless}. Recent studies~\cite{ramakrishnan2016reconstruction} indicate that EEG channels exhibit high cross-correlation, known as \emph{functional connectivity}. Leveraging this insight and motivated by~\cite{antar2023functional, price2021attention}, we employ a channel-wise functional connectivity-based approach to reconstruct \emph{enhanced} EEG signals from low-density devices, improving their ability to capture high-attention episodes related to cognitive and ERP events. To synchronize EEG data with other modalities, we downsample the sampling frequency from 128 Hz to 32 Hz using a decimation approach. The decimation factor is computed as:  $ \frac{Org_{fs}}{Tar_{fs}} $, where $Org_{fs}$ and $Tar_{fs}$ denote the original (128 Hz) and target (32 Hz) sampling frequencies, respectively. To mitigate aliasing, a Butterworth low-pass filter with a cutoff frequency of $0.5 \times Tar_{fs}$ Hz (14 Hz) is applied before decimation, preventing temporal shifts and signal distortion. The cutoff frequency is normalized using $ \frac{Tar_{fs}}{Org_{fs}} $, and a filter order of 4 is employed, assuming the EEG signal is band-limited below the Nyquist frequency.

\begin{figure*}
\centering
    \begin{minipage}{0.48\textwidth}
        \centering
        \includegraphics[width=2.8in,height=1.5in]{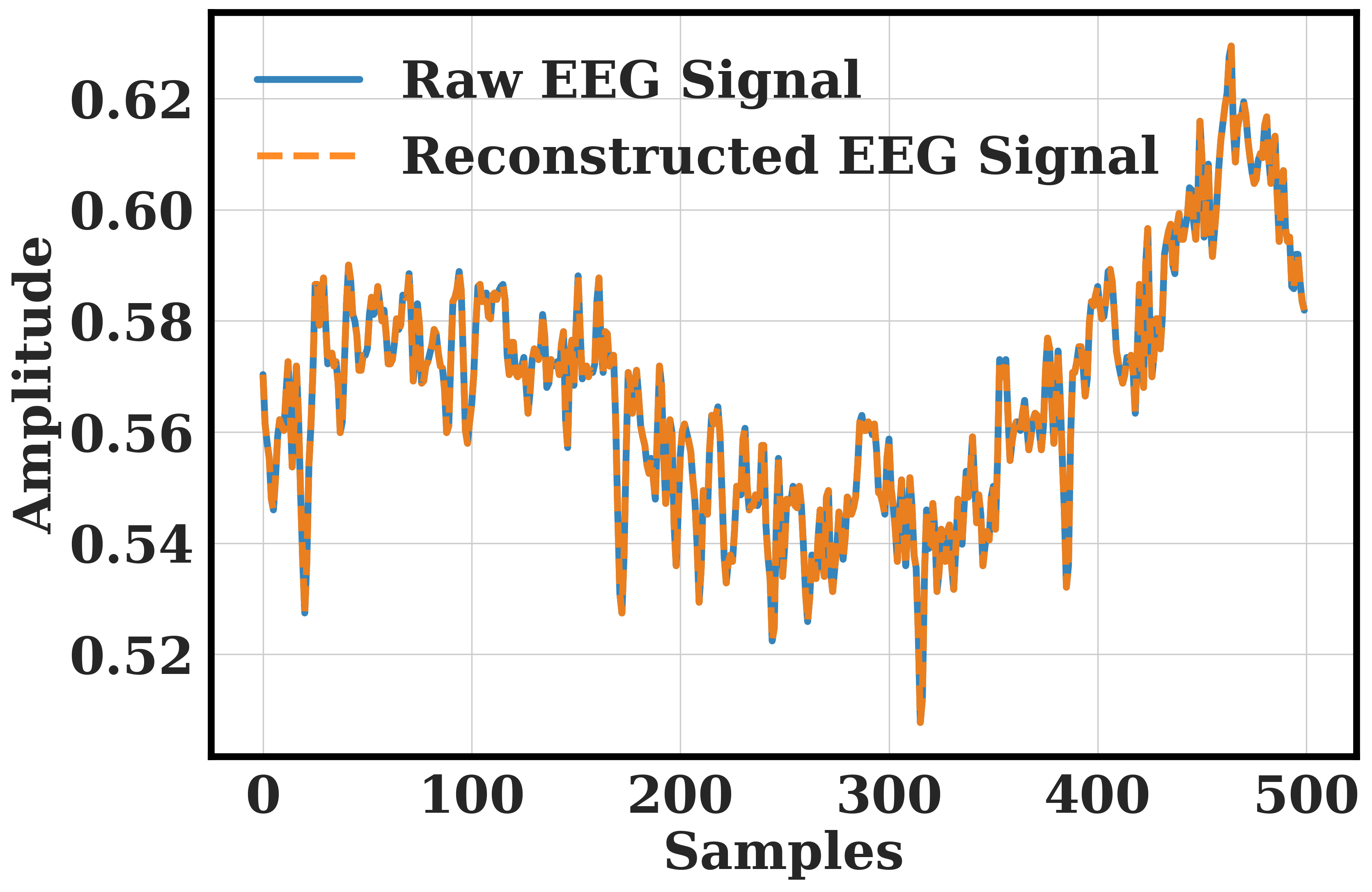}
        \subcaption{Original EEG channel against the reconstructed signals.}
        \label{fig:recon_ori_signal}
    \end{minipage}%
    \hfill 
    \begin{minipage}{0.515\textwidth}
        \centering
        \includegraphics[width=2.8in,height=1.5in]{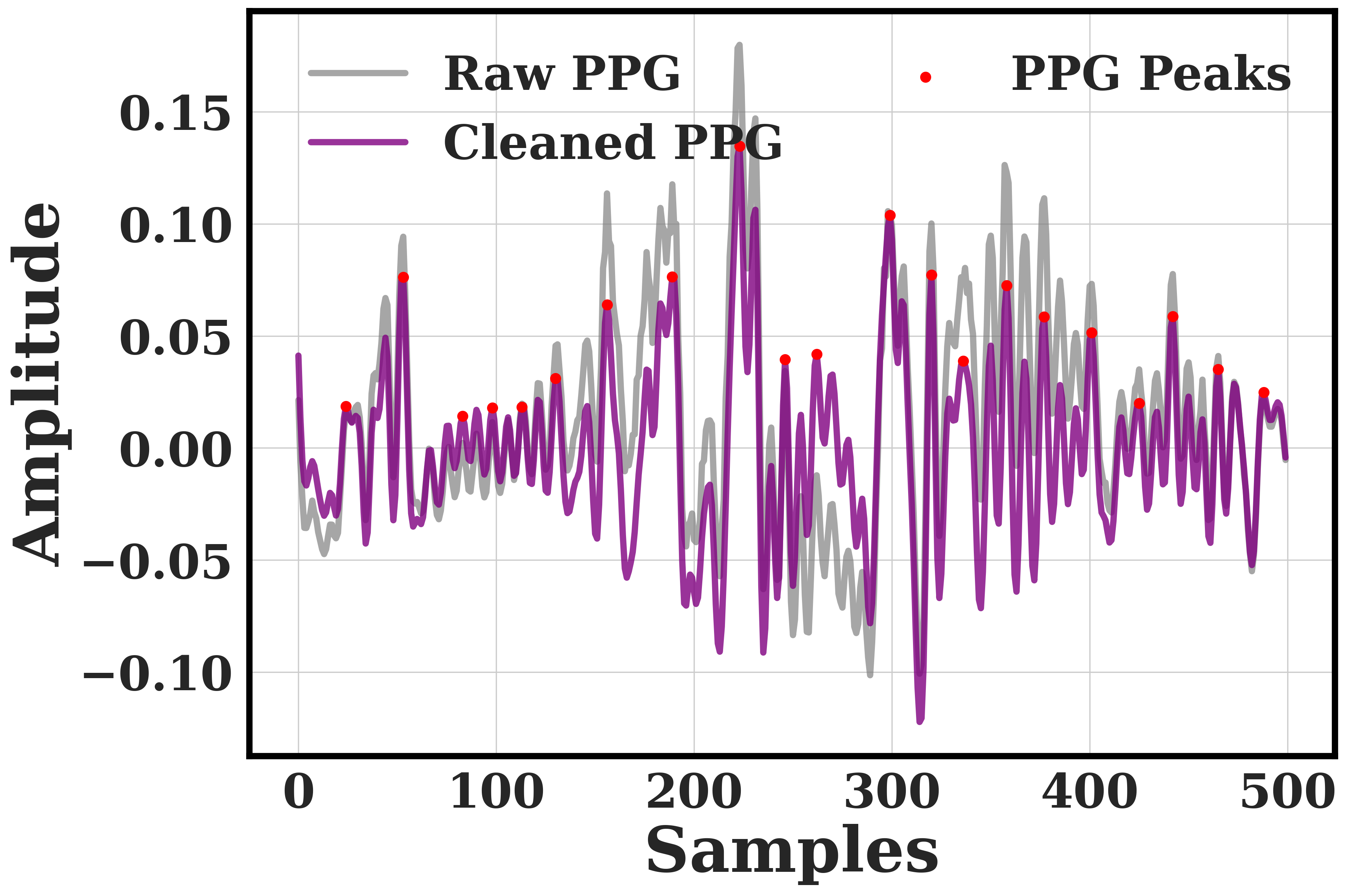}
        \subcaption{Raw and cleaned PPG signals with peaks corresponding to heartbeats}
        \label{fig:ppg_signal}
    \end{minipage}
      \caption{Comparative analysis highlighting the effectiveness of signal processing for GSR and PPG modalities for better understanding the events/states changes during the visual stimuli.}
\end{figure*}

\textbf{Reconstruction:} We perform Independent Component Analysis (ICA)~\cite{stone2002independent}, commonly used for Blind Source Separation (BSS), on the filtered signals, assuming high-density electrode availability during training. ICA separates incoming signals into their respective sources by assuming statistical independence and non-Gaussianity. To evaluate the reconstructed signals, we compute the cosine similarity~\cite{6410152, niknazar2018new} (Eq.~\ref{eq:cosine}) between the transformed ICA matrix ($t$) and the subset of EEG signals ($e$) from the low-density device. The reconstruction effectiveness is demonstrated in Figure~\ref{fig:recon_ori_signal}, highlighting \name's capability to separate high-variability noise and preserve the underlying temporal representations of EEG signals.

\begin{align}
\begin{split}
& \text{cosine similarity} ({\bf t},{\bf e})= \frac{{\bf t} \cdot {\bf e}}{\|{\bf t}\| \|{\bf e}\|} = \frac{ \sum_{i=1}^{n}{{\bf t}_i{\bf e}_i} }{ \sqrt{\sum_{i=1}^{n}{({\bf t}_i)^2}} \sqrt{\sum_{i=1}^{n}{({\bf e}_i)^2}}} \\ 
& \hspace{4em} \text{cosine distance} = 1 - \text{cosine similarity}
\label{eq:cosine}
\end{split}
\end{align}

\noindent \textbf{Galvanic Skin Response (GSR, $\mathcal{G}^{R,C}$).} GSR measures changes in skin conductance, providing insights into autonomic nervous system activity and emotional arousal, which correlate with cognitive performance deterioration~\cite{posada2017sleep, posada2019phasic}. Motivated by this, we extract the \textit{phasic} component of the Skin Conductance (SC) signal for downstream event detection. First, we remove noise by applying a \textit{Butterworth band-pass filter} (0.1–5 Hz). Then, we employ the CvxEDA algorithm~\cite{greco2015cvxeda}, a convex optimization method, to decompose the SC signal into \textbf{phasic}, \textbf{tonic}, and \textbf{white noise} components. The phasic component captures fluctuations linked to cognitive and emotional states (e.g., arousal, cognitive workload)~\cite{bradley2000affective, gashi2020detection}. As illustrated in Fig.~\ref{fig:gsr_signal}, \textit{Skin Conductance Response (SCR) peaks} mark heightened activation phases, while \textit{SCR onsets} indicate the initial rise from baseline. The \textbf{tonic component} represents the baseline skin conductance level, whereas the \textbf{white noise component} accounts for measurement artifacts. Additionally, \textit{SCR half-recovery} marks the point where skin conductance has decreased halfway between the peak and baseline, reflecting the recovery phase post-arousal.

\noindent \textbf{Photoplethysmograph (PPG, $\mathcal{P}^{R,C}$).} PPG enables the extraction of key physiological parameters, including heart rate, blood pressure, and respiratory rate~\cite{lei2020estimation}. The PPG signal reflects pulsatile blood volume changes in response to incident light, with \textit{peaks} and \textit{valleys} corresponding to distinct cardiovascular events. The waveform consists of the \textbf{systolic peak}, representing maximal blood volume during heart contraction, and the \textbf{diastolic notch}, indicating minimal blood volume during relaxation. Prior studies have shown that PPG can effectively diagnose acute stress in high-stress environments~\cite{wright2014increased}. Motivated by this, we adopt the approach in~\cite{elgendi2013systolic} to extract time-series features such as heart rate and PPG peaks/valleys. The goal is to estimate \textbf{systolic blood pressure (SBP)}, which is known to be influenced by visuospatial cognitive tasks~\cite{swan1998systolic}. To extract peaks, we apply a \textbf{Chebyshev-II 4$^{th}$ order band-pass filter} (0.1–5 Hz) to remove noise, following~\cite{liang2018optimal}. Figure~\ref{fig:ppg_signal} compares raw and filtered PPG signals, where detected peaks reflect blood pressure variations during stimuli, providing a refined representation of pulse waveforms.

\subsection{Multimodal Fusion}
\label{sec:augmentation}

\name integrates multimodal physiological data (EEG, GSR, and PPG) to enhance the detection of high cognitive load or attentional focus episodes. Our user studies utilize ELAN\footnote{https://archive.mpi.nl/tla/elan} to synchronize data from multiple devices for time alignment. Participants are instructed to perform a "NOD" (bending their head up and down) and "WAVE" their dominant hand three times at the start and end of each task to mark timestamps. While automated synchronization methods exist in the literature~\cite{bannach2009automatic}, our focus remains on evaluating the effectiveness of multimodal wearable sensors in extracting cognitive load and their practical usability.

\begin{equation}
    \mathbf{{\it Fus}}_{\theta}^{R,C} = [\mathcal{E}_{\theta}, \mathcal{G}_{\theta}, \mathcal{P}_{\theta}]^{R,C}
    \label{eq:fus}
\end{equation}

We employ an \textit{incremental} feature-level fusion approach~\cite{chen2015feature} to integrate EEG, GSR, and PPG signals for ERP detection during stimuli. Features from each modality are independently extracted, preprocessed, and standardized using Min-Max scaling to improve signal quality. The final multimodal representation is obtained by aggregating the normalized features across modalities, as defined in Eq.~\ref{eq:fus}. Synchronized gaze data collected during navigation tasks further demonstrate the effectiveness of GSR and PPG in capturing ERP episodes (see Section~\ref{sec:preprocessing}).

\subsection{Event-related Potentials~(ERPs) Episodes Extraction}
\label{sec:cpd}

Next, we apply change point detection (CPD)~\cite{aminikhanghahi2017survey} to identify abrupt changes in the time series data when a property of the signal shifts. We adopt a sliding window-based CPD approach for \name due to its lower complexity and competitive performance on similar time series tasks~\cite{silva2021time} (see Section~\ref{sec:discussion} for runtime performance). Mathematically, sliding window CPD is defined using a cost function \( C(y)_i = \sum_{t=i} ||y_{i} - \overline{y_{i}}||_{1} \), where \( ({y_t})_t \) represents the input signal, and the discrepancy function
\( d(y_{u,v}, y_{v,w}) = C(y_{u,w}) - C(y_{u,v}) - C(y_{v,w}) \) measures the cost gain when splitting the subsignal \( y_{u...w} \) at the index \( v \). If the statistical properties of the sliding windows \( u...v \) and \( v...w \) are similar, the discrepancy is low; otherwise, a significant discrepancy indicates a change point. Sliding window CPD is widely used for non-stationary data to detect abrupt changes~\cite{naqvi2020real, chu1995time}. However, EEG signals are prone to artifacts such as muscle movements, blinks of the eyes, electrode interference, and environmental noise, which can distort the signal and affect the reliability of the analysis~\cite{sadiya2021artifact}.

To mitigate artifacts and effectively capture ERP episodes, we employ Continuous Wavelet Transform (CWT), which decomposes EEG signals into wavelet functions for joint time-frequency analysis. CWT is particularly effective in detecting transient events such as ERPs and identifying artifacts localized within specific time segments~\cite{kumar2021classification, qassim2013wavelet}. We utilize the Morlet wavelet for its optimal time-frequency localization, ability to capture oscillatory components, superior frequency selectivity, adaptability to nonstationary signals, and proven utility in ERP detection, with the CWT defined as: \(\frac{1}{\sqrt{|a|}} \int_{-\infty}^{\infty} {Fus}_{\theta}^{R,C} (t) \psi^*\left(\frac{t - b}{a}\right) dt \).

The ERP extraction approach integrates continuous wavelet transform (CWT) with sliding window CPD, as defined in Eq.~\ref{eq:wincwt}. Here, \( t_i = i \cdot \Delta t \) represents the start time, \(\Delta t\) is the step size, \(\text{CWT}_i(a, b)\) denotes the CWT of the \(i\)-th window, \(w\) is the window length, and \(x(t)\) is the signal. This method leverages sliding window's localized analysis and wavelet transformation's multi-resolution capabilities for EEG signal reconstruction. A grid search optimized the window size for ERP extraction, exploring durations from 0.1 to 1.0 seconds, with an optimal window size of \textbf{\textit{0.75}} seconds. Due to varying response delays across physiological modalities, resulting in change points or peaks occurring at different times, the sliding window approach effectively addressed this challenge.

\begin{equation}
\begin{aligned}
{Fus}_{\theta, i}^{R,C}(t) &= {Fus}_{\theta}^{R,C}(t + t_i), \quad t \in [t_i, t_i + w]\\
{Memento}^{CWT+CPD}_i(a, b) &= \frac{1}{\sqrt{|a|}} \int_{0}^{w} {Fus}_{\theta, i}^{R,C}(t) \psi^*\left(\frac{t - b}{a}\right) dt
\end{aligned}
\label{eq:wincwt}
\end{equation}

\section{Working Memory Dataset} 
Our in-house dataset comprises \textbf{two user studies} designed to address \textbf{RQ1} and \textbf{RQ2} outlined in the introduction. \textbf{Study 1} focuses on extracting high-attentional ERP episodes from multimodal physiological signals (EEG, GSR, and PPG), while \textbf{Study 2} examines the effectiveness of these ERPs as intelligent cues to enhance navigation recall and personalized short-term memory. Additionally, we employed lightweight wearable sensors, including the Emotiv Insight headset (5 electrodes) and a single Shimmer device (PPG + GSR) worn on the dominant hand to ensure minimal user burden during data collection. A 5-point self-reported comfort scale is used to assess feasibility, with 76.4\% of participants rating the setup as highly comfortable and minimally intrusive, indicating the practicality of the sensor configuration.

\footnotetext[1]{We have released the in-house \textbf{\textit{\href{https://sites.google.com/umbc.edu/wmdataset/home}{https://sites.google.com/umbc.edu/wmdataset/home}}} (WoM dataset), along with the implementation, to support reproducibility and engagement within the research community.}

\subsection{Study 1: High-attentional ERP Episodes Extraction via Physiological Signals}\label{sec:study1}

A total of {\bf 15} participants (7 male, 8 female), aged 18–29, participated in this \textbf{IRB-approved} study. Participants were recruited through institutional email solicitations without monetary compensation. Individuals with medical implants, such as pacemakers, were excluded for safety reasons. Written informed consent was obtained following a study briefing. The study aimed to evaluate \textbf{(a)} the effectiveness of \namef in extracting meaningful change points during a search and navigation task and \textbf{(b)} Validation of findings that \emph{cueing} enhances memory recall.

\subsubsection{Study Environments:}\label{sec:env}  
\begin{figure}
\begin{center}
  \begin{minipage}[b]{.22\columnwidth}
    \includegraphics[scale =0.045]{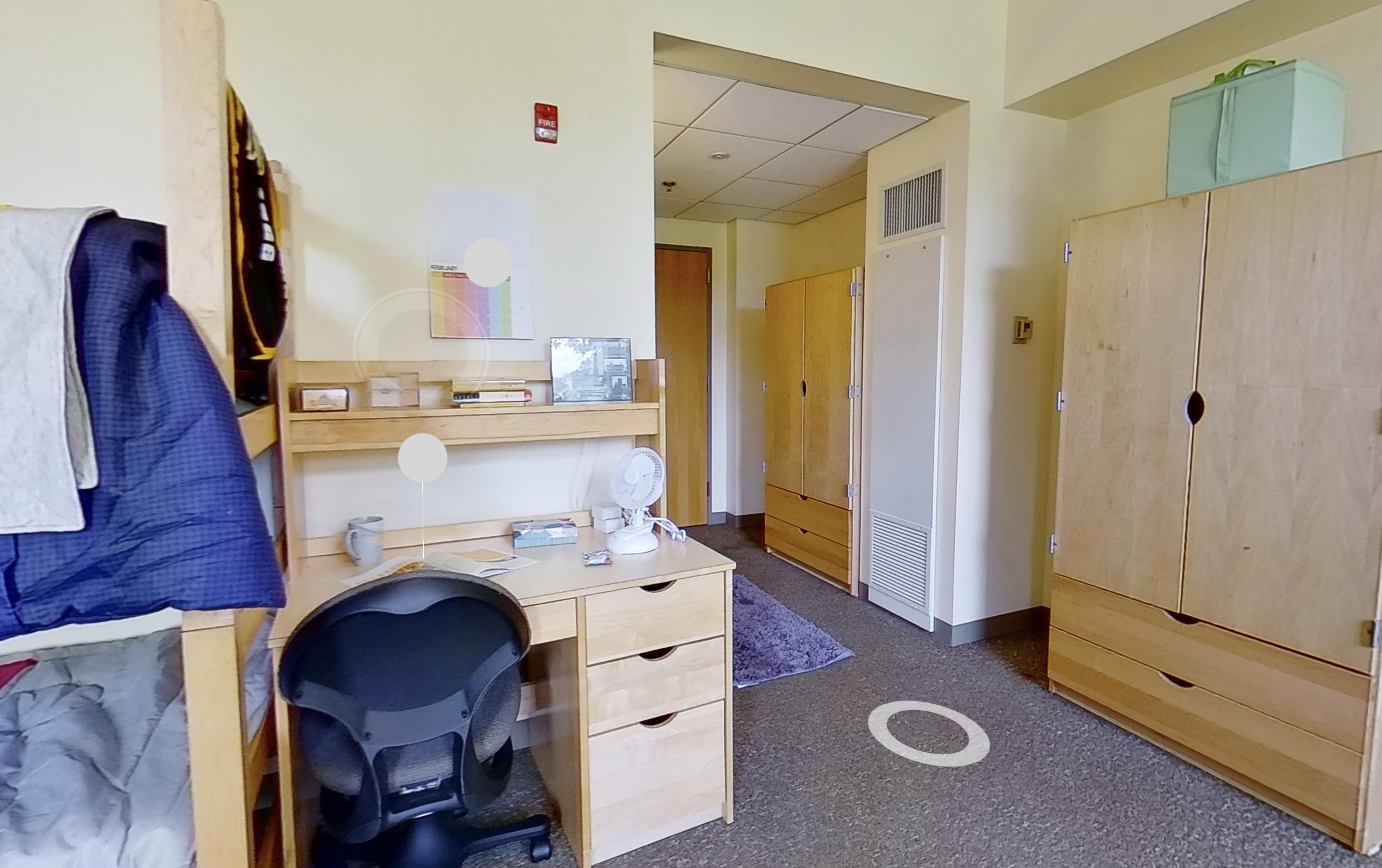}
    \subcaption{Indoor}
  \end{minipage}
\hfill
    \begin{minipage}[b]{.22\columnwidth}
    \includegraphics[scale =0.045]{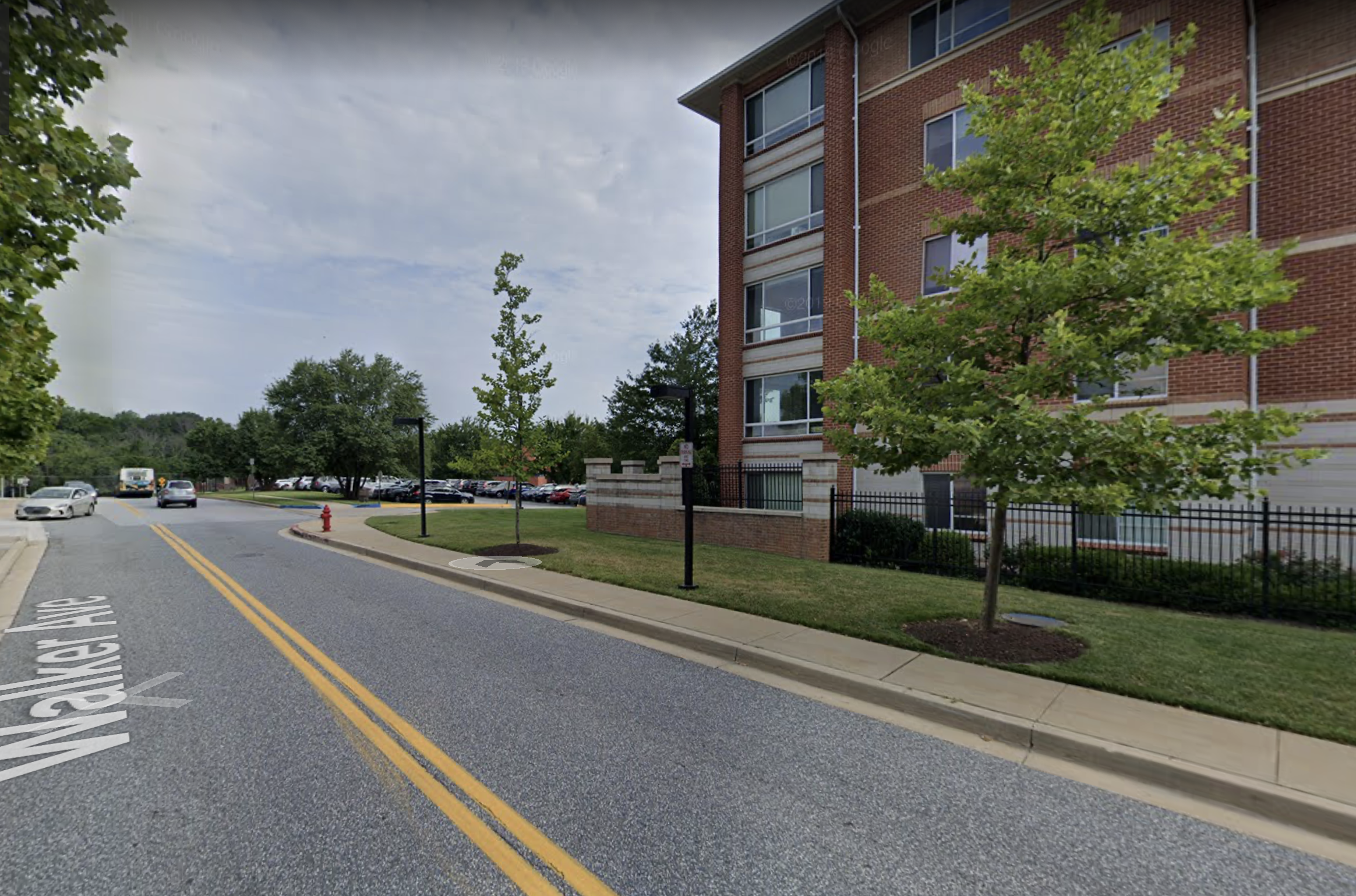}
    \subcaption{Campus}
    \end{minipage}
\hfill
    \begin{minipage}[b]{.22\columnwidth}
    \includegraphics[scale =0.045]{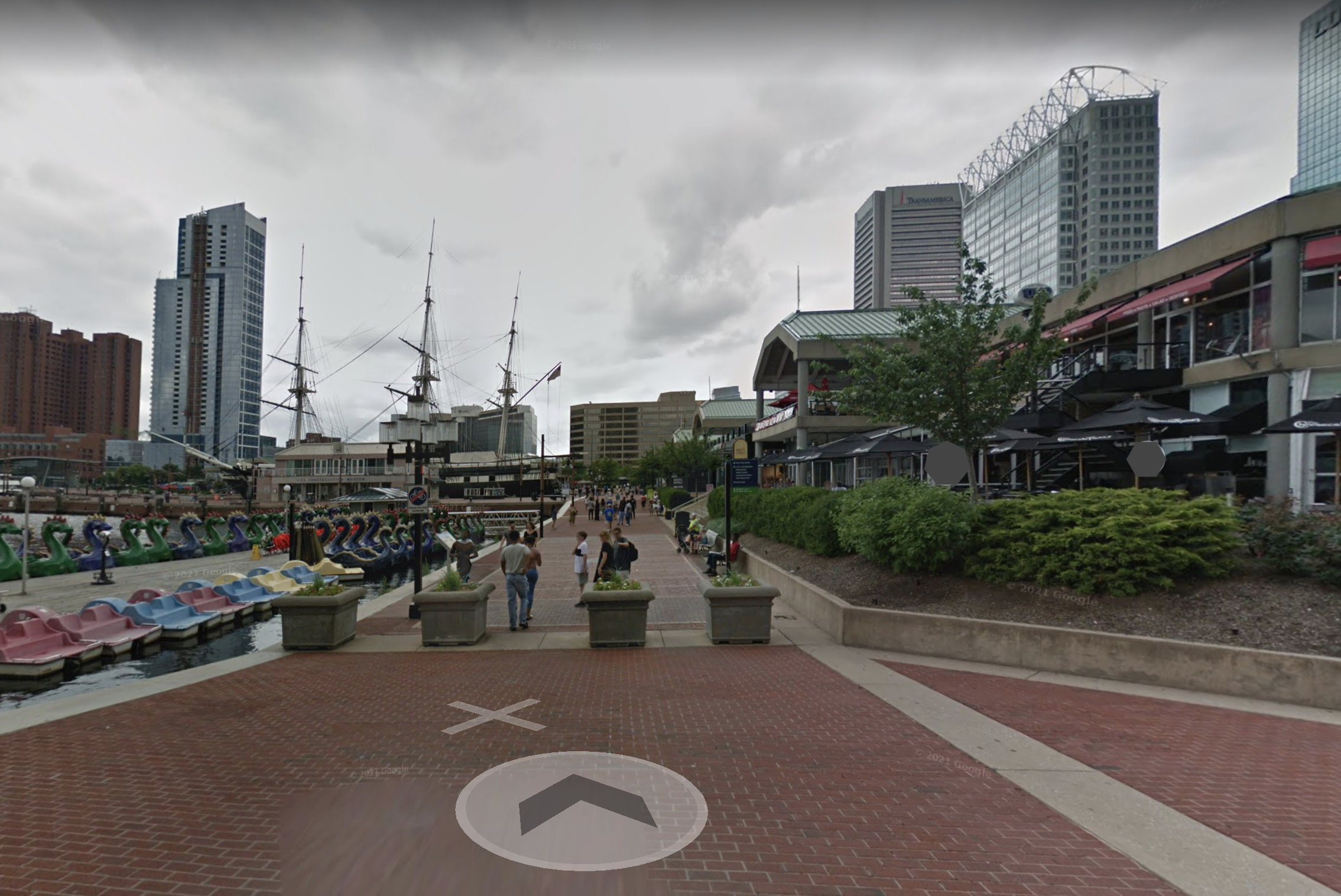}
    \subcaption{Downtown Baltimore}
    \end{minipage}
\hfill
    \begin{minipage}[b]{.22\columnwidth}
    \includegraphics[scale =0.045]{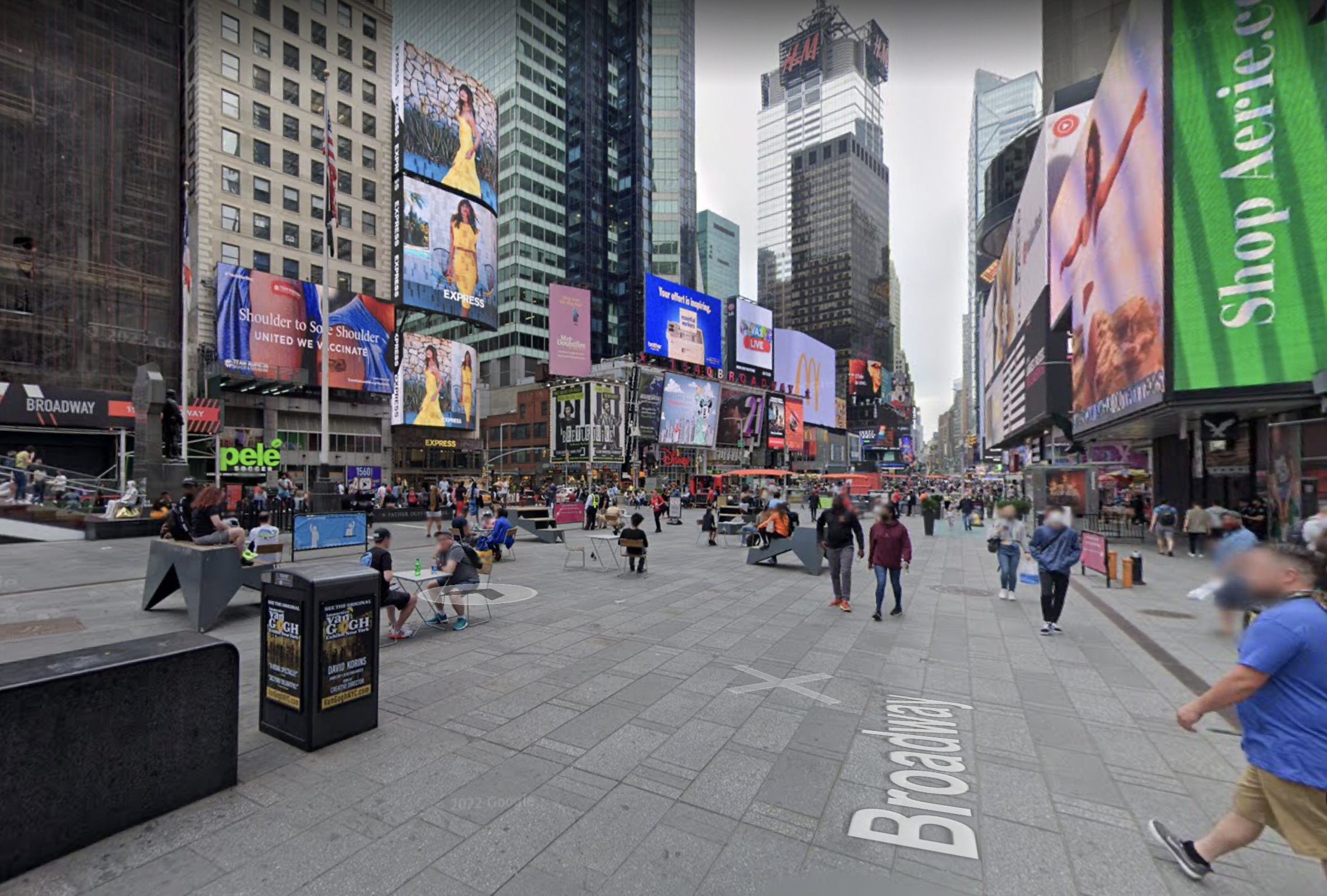}
    \subcaption{NYC}
    \end{minipage}
    \end{center}
    \caption{Virtual Environments used in the Studies} 
    \label{fig:env_simulated}
    \end{figure}

Similar to prior work in wayfinding and route tracing~\cite{lingwood2018using, ye2022cognitive}, we employed a \textit{desktop-based virtual environment} where participants navigated through four distinct scenarios: \textbf{(a)} an indoor dorm, \textbf{(b)} a familiar suburban campus area, \textbf{(c)} the downtown area of a mid-size U.S. city (Baltimore), and \textbf{(d)} a dense cosmopolitan city (NYC). Figure~\ref{fig:env_simulated} presents sample snapshots of these environments, highlighting their varying complexity. Participants wore Emotiv Insight headsets equipped with EEG and IMU sensors, along with Shimmer sensors for GSR and PPG data collection. Throughout the navigation tasks, screen recordings and gaze fixations were captured using the freely available Gaze Recorder\footnote{\url{https://gazerecorder.com}}, which provided ground truth on participants' visual attention and detected artifacts such as eye blinks. The Gaze Recorder features a user-friendly gaze calibration interface consisting of two steps: \textbf{(i)} Facial synchronization, where participants align their faces by following red points on the screen, and \textbf{(ii)} Gaze calibration, where participants track {\it 16 points} appearing across the screen. During task performance, fixation heatmaps were generated for areas of interest (AOI). Task start and end times were recorded to synchronize multimodal data streams from the Shimmer and Emotiv Insight headsets, and each session was recorded using an Akaso action camera\footnote{\url{https://www.akasotech.com/ek7000}}, serving as a ground truth reference for further analysis.

\begin{figure}
    \centering
    \includegraphics[scale =0.375]{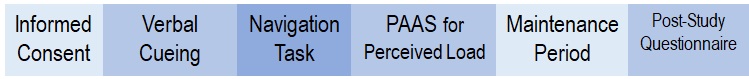}
  \caption{Protocol followed for Study 1}
    \label{fig:study1-protocol}
\end{figure}

\subsubsection{Study Protocol:}  

This study aims to: \textbf{(a)} Evaluate whether EEG reconstruction and multimodal sensor integration in \name enhance the efficacy of high-attentional event extraction and \textbf{(b)} Validate prior findings~\cite{ye2016retro, heuer2016feature, olivers2006feature} on the effectiveness of \emph{cueing} in improving working memory recall. The study protocol, illustrated in Figure~\ref{fig:study1-protocol}, was followed for each environment. Participants were \emph{verbally cued} to focus on specific objects during the navigation task, starting from a common initial point (see Table~\ref{tab:cued} for cued objects). The navigation task duration was limited to \textbf{1 minute}, ensuring consistency across trials by keeping participants within the same areas. Following a \textbf{5-minute maintenance period}, participants completed the PAAS scale~\cite{paas1994measurement} and were asked to \emph{recall} the number of encountered objects, including both cued and non-cued objects. During the maintenance period, participants were instructed to relax, and no additional cognitive tasks were assigned.

\begin{table}
\caption{Verbally Cued and Non-Cued Objects in Study 1}
\scalebox{0.74}{
\begin{tabular}{|l|c|c|}
\hline {\bf Environment}        & {\bf Cued Objects}  & {\bf Non-Cued Objects~(Questionnaire)} \\ \hline
Indoor             & Table, Laptop & Bed, Chair                            \\
Campus             & Bus, Truck     & Car                                   \\
Downtown Baltimore & Boat          & Bench                                 \\
New York City      & Billboard     & Bicycle      \\ \hline                        
\end{tabular}}
\label{tab:cued}
\end{table}

\subsection{STUDY 2: Improving of Navigation Memory Recall through ERP Episodes Cueing}\label{sec:study2}
\begin{figure}
   \includegraphics[scale=0.28]{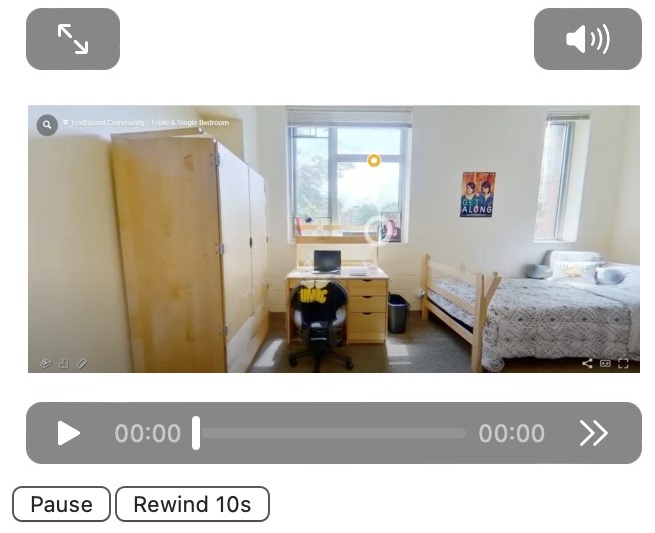}
 \caption{Participant View of the Navigation Highlighter Tool.}
     \label{fig:summary-tool}
   \end{figure}

In this second study, we assess the feasibility of utilizing physiological sensing to augment short-term memory. Similar to Study 1, we focus on a navigation task where participants retrace a previously explored route to evaluate their \emph{recall} of the episode.  We first present the implementation details of the highlighting tool, as shown in Fig.~\ref{fig:summary-tool}. Next, we outline the study protocol (Section~\ref{sec:study2pro}). Finally, we provide a quantitative analysis of the effectiveness of physiological-driven and computer vision-based cueing on recall performance (Section~\ref{sec:s2-recall}).

\begin{figure}
    \centering
    \includegraphics[scale =0.3]{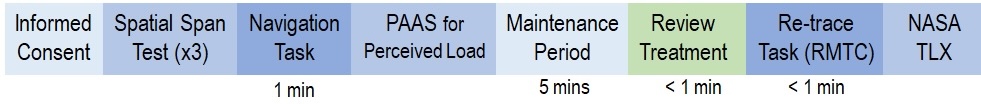}
    \caption{Protocol followed for Study 2} 
    \label{fig:study2-protocol}
\end{figure}

\subsubsection{Study Protocol:}~\label{sec:study2pro} A total of \textbf{25 participants} (15 male, 10 female), aged 19--32, took part in this \textbf{IRB-approved} study, following the protocol illustrated in Figure~\ref{fig:study2-protocol}. Participants were distinct from those in Study 1. After obtaining written informed consent, participants completed the Spatial Span Test~\cite{woods2016improved}, a freely available online assessment\footnote{\url{https://www.memorylosstest.com/visual-span/}}. Since visuospatial working memory capacity can influence recall performance, this test was administered to control for variability. Each participant repeated the test three times with a fixed span length of 6, and their average correct sequence length was recorded as a measure of visuospatial memory capacity.  

Each participant performed \textbf{three} navigation tasks—\textit{indoor}, \textit{campus}, and \textit{mid-size city}—in a randomized order to mitigate ordering effects. Each task lasted \textbf{5 minutes}, followed by a PAAS-based cognitive load self-assessment~\cite{paas1994measurement}. Similar to Study 1, participants underwent a \textbf{5-minute} maintenance period with no additional cognitive tasks. Participants then received one of the following \textit{evenly distributed} treatments: \textbf{(i)} \textit{No Aid (Free Recall)}—recall without visual cues, \textbf{(ii)} \textit{All Frames}—full navigation replay, \textbf{(iii)} \textit{Random}—random frame selection, \textbf{(iv)} \textit{CV}—CV-selected memorable frames, \textbf{(v)} \textit{CV-EEG}—CV-selected frames filtered by EEG change points, \textbf{(vi)} \textit{CV-All}—CV-selected frames filtered using all physiological signals (EEG, GSR, and PPG), \textbf{(vii)} \textit{Memento-EEG}—replayed frames identified as memorable via EEG, and \textbf{(viii)} \textit{Memento-EGP}—replayed frames identified by \name using all physiological signals.  

Computer vision-based approaches employ the state-of-the-art memorability estimation model, AMNet~\cite{fajtl2018amnet}. The memorability Pearson coefficient threshold is set to \textbf{0.677} for all experiments, as established in the original study. We utilize the pre-trained weights of the $ResNet101FC$ model trained on the LaMem dataset to ensure robust CV-based memorability predictions. During the \textit{review treatment phase}, participants interacted with the highlighter tool for \textbf{one minute}, except in the Free Recall condition. For treatment cases \textbf{(2)} through \textbf{(8)}, participants completed the NASA TLX scale~\cite{hart1988development} to assess the cognitive load induced by each treatment. After the treatment, participants were asked to retrace their route in the virtual environment within the same \textbf{1-minute} window, following a protocol similar to the Route Map Recall Test (RMRT)~\cite{wang2012validation}. Upon completion, participants provided qualitative feedback through an open-ended questionnaire, documenting significant landmarks and objects encountered during the navigation task.

\begin{figure*}[t]
\begin{minipage}{\textwidth}
\centering
        \begin{subfigure}[t]{0.45\textwidth}
	   \centering \includegraphics[scale= 0.155]
{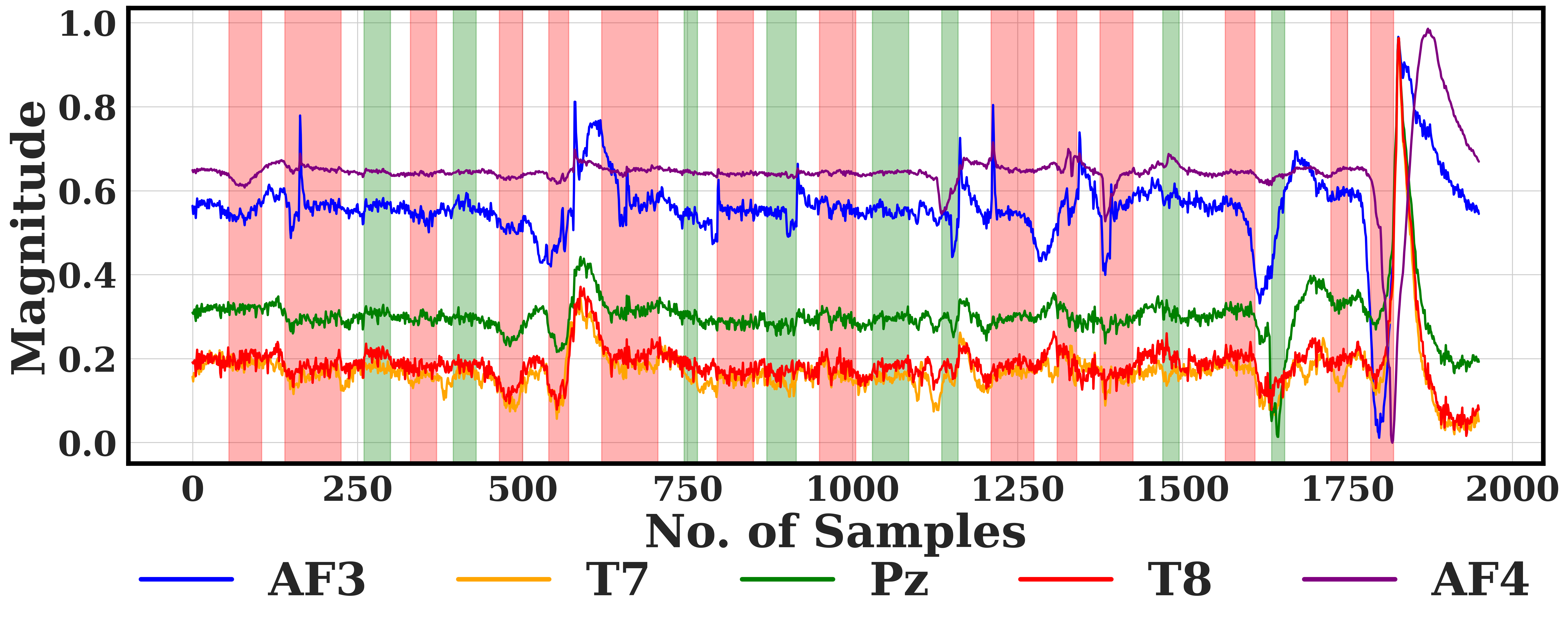}
      \subcaption{Raw EEG Signals}
        \end{subfigure}%
       \hfill%
        \begin{subfigure}[t]{0.55\textwidth}
	   \centering \includegraphics[width=3.9in,height=1.21in]{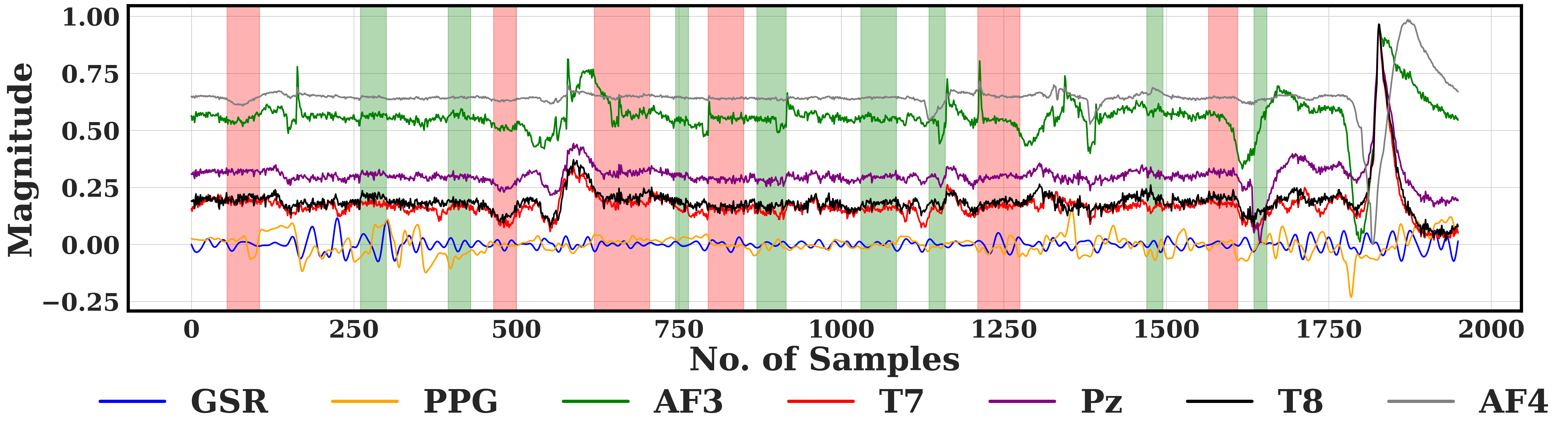}
        \subcaption{All modality (EEG, GSR and PPG) signals.}
        \end{subfigure}%
    \caption{Comparative analysis of effectiveness in detecting High attentional ERP episodes for two analyses: (a) {\it Raw EEG Signals} and (b) {\it All Modal Signals}, respectively.}
     \label{fig:CWT_com_modal}
\end{minipage}
\end{figure*}

\section{Results and Discussion}~\label{sec:result}

Here, we highlight the key findings from our experiments, analyzing the effectiveness of \name in enhancing navigation memory recall through physiological signals. We present quantitative results comparing the performance of different treatment conditions, including free recall, CV-based, and physiological-based approaches. 

\subsection{RQ1: Extraction of High-Attentional ERPs}  
We first evaluate the effectiveness of multimodal sensing in extracting attentional episodes using verbal cues to \emph{direct attention}. Specifically, we (i) compare the CWT approach against change point detection (CPD) alone, and (ii) demonstrate that participants recall \emph{cued objects} more effectively.

\noindent {\bf {\underline{Accuracy of Change-Point Detection to extract ERPs:}}}  
To assess the effectiveness of \names, we used auxiliary gaze input from the recorder application as ground truth for visual attention. The application generates gaze heatmaps, where prolonged dwell times result in \emph{redder} and \emph{wider} fixation points, providing a visual representation of attention intensity.

\begin{figure*}[t]
\centering
\begin{subfigure}[t]{0.46\textwidth}
    \centering 
    \includegraphics[scale= 0.178]{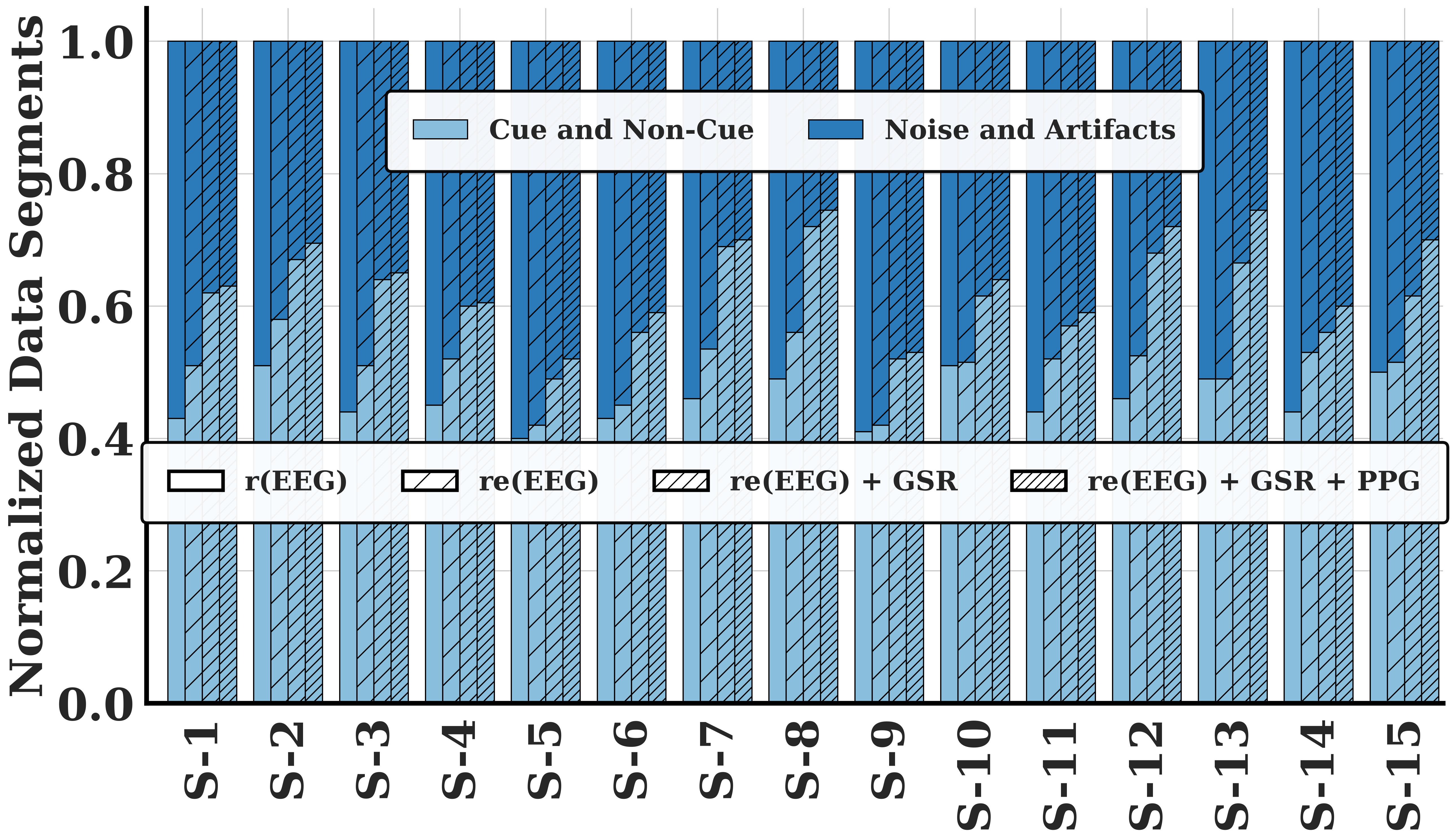}      \caption{The proportion of change point segments detected during the indoor navigation task using (a) raw EEG signals, (b) reconstruct EEG signals, (d) reconstruct EEG + GSR + PPG signals, for each participant, where {\bf r} and {\bf re} corresponds to \underline{r}aw and \underline{re}constructed signals, respectively across all participants.}
      \label{fig:cpd_non_cued_cued_segments}
\end{subfigure}%
\hfill
\begin{subfigure}[t]{0.24\textwidth}
     \centering \includegraphics[scale= 0.18]{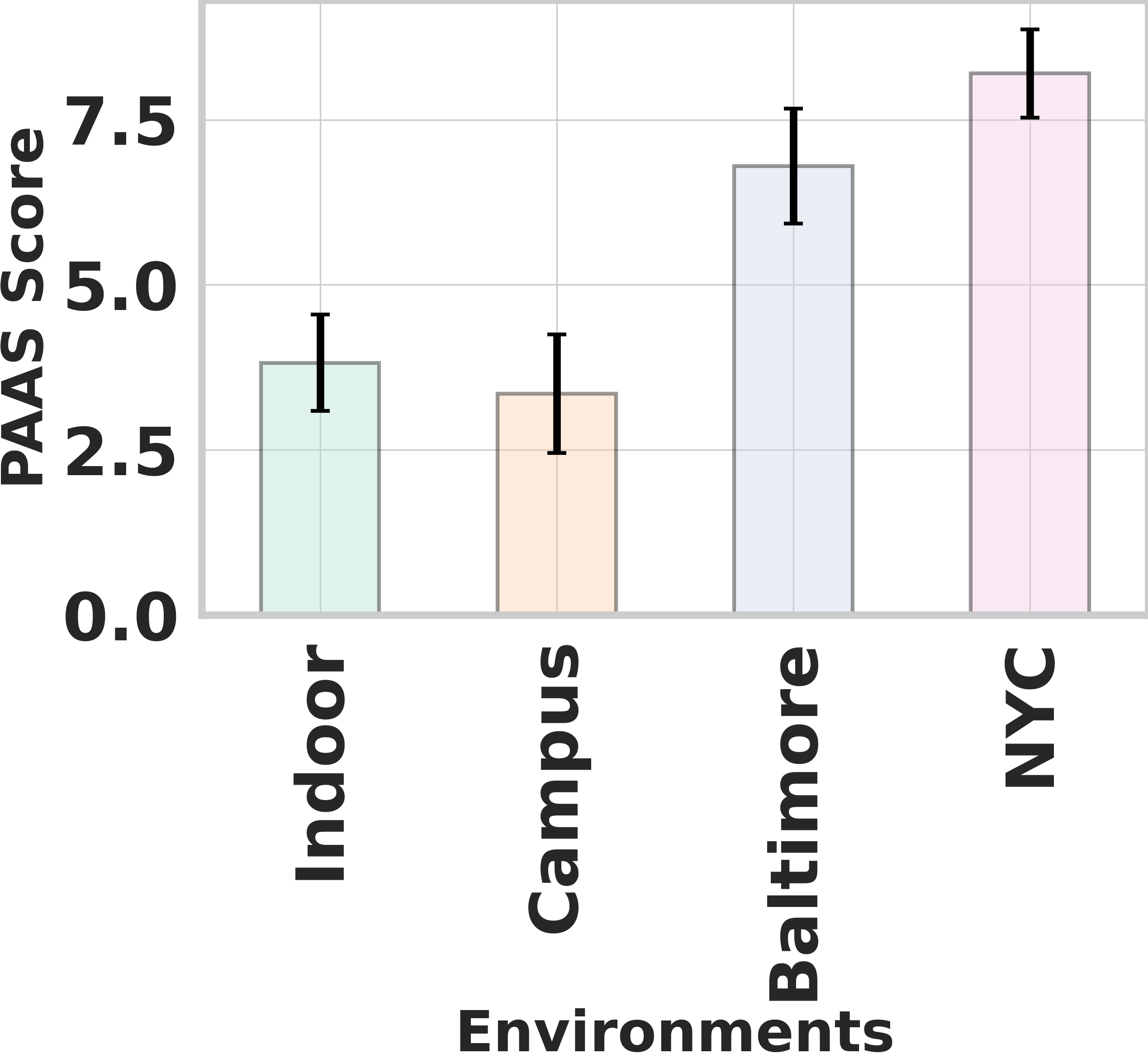}
     \caption{Average PAAS score based on each participant's cognitive load during each environment.}
     \label{fig:pass_env}
\end{subfigure}%
\hfill
\begin{subfigure}[t]{0.275\textwidth}
    \centering \includegraphics[width=1.81in,height=1.47in]{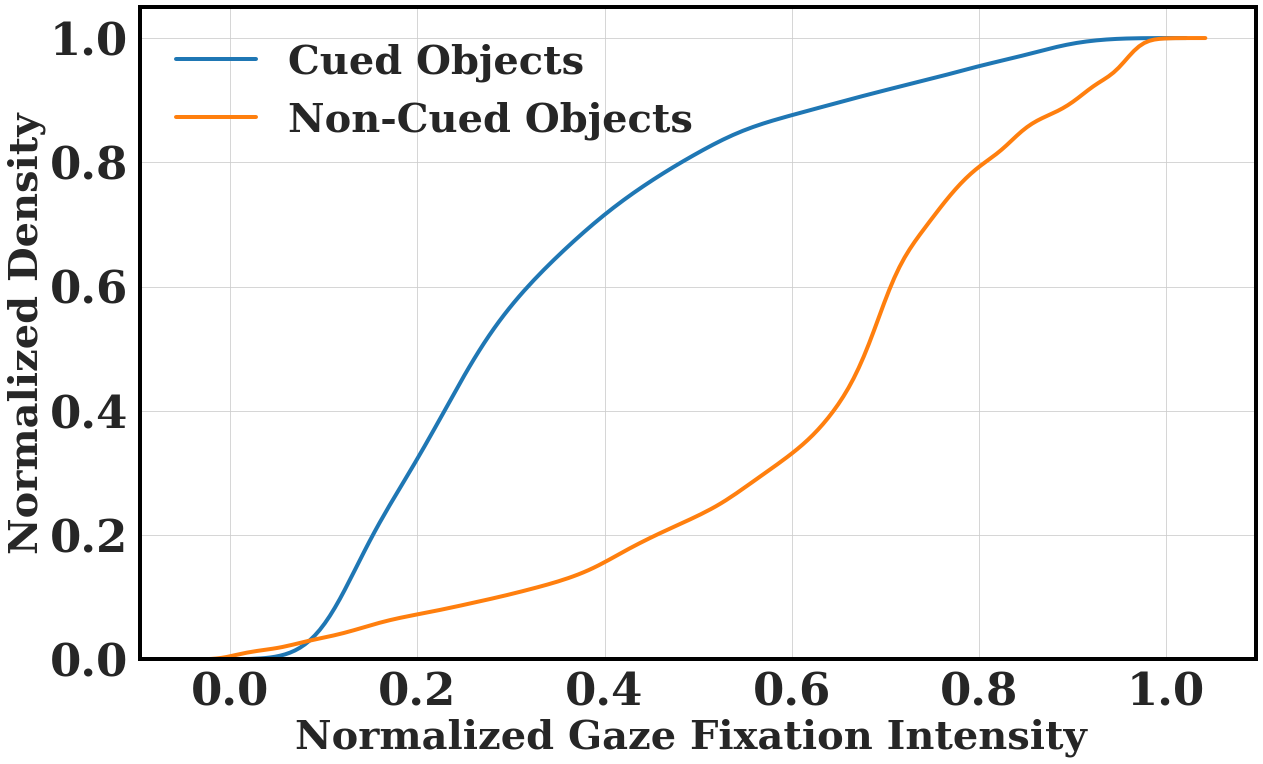}
    \caption{Cumulative Density plot for Cued and Non-Cued Objects (listed in the questionnaire).}
     \label{fig:CDFplot}
\end{subfigure}
\caption{Results for study 1 for episode extraction and verbal cueing task.}
\label{fig:Result_study_1}
\end{figure*}

\noindent In Figure~\ref{fig:cpd_non_cued_cued_segments}, we present the percentage of \emph{change points} using $\frac{X_{i}}{X}$, where $X_{i}$ represents fixation instances or visual attention periods toward cued/non-cued objects during visuospatial working memory tasks, and $X$ denotes the total number of attention-related change points, including artifacts. The analysis spans (a) raw EEG, (b) reconstructed EEG, (c) reconstructed EEG + GSR, and (d) reconstructed EEG + GSR + PPG signals. The results indicate a significant improvement in detecting neural-response spikes corresponding to \emph{attentional instances}, increasing from \textbf{49.8\%} to \textit{72.6\%} across all modalities using the \name framework. Noise-affected instances (e.g., eye blinks and random movements, manually labeled by the experimenter) were evaluated, revealing that \name detects an average of \textbf{1920} change points per user, with approximately \textbf{1498} instances (standard deviation of \textbf{40}) classified as \emph{valid}, capturing attentional periods related to both cued and non-cued objects. The proportion of noisy detections across modalities was: (a) raw EEG: {\bf 50.2\%}, (b) reconstructed EEG (Recon EEG): {\bf 39.6\%}, (c) reconstructed EEG + GSR: {\bf 30.8\%}, and (d) reconstructed EEG + GSR + PPG: {\bf 27.4\%}, as summarized in Table~\ref{tab:modal_fusion} and Fig.~\ref{fig:Result_study_1}. The inclusion of reconstructed EEG + GSR + PPG signals resulted in a {\bf 22.8\%} reduction in noise and artifacts, significantly enhancing ERP episode extraction robustness. Furthermore, raw EEG signals yielded \textbf{22} detected segments, including \textbf{8} ERP episodes and \textbf{14} noise artifacts. With multimodal fusion, the total detected segments decreased to \textbf{14}, preserving \textbf{8} ERP episodes while reducing noise artifacts to \textbf{6}, as shown in Fig.~\ref{fig:CWT_com_modal}, where \textbf{RED} and \textbf{GREEN} indicate noise and ERP segments, respectively. These results demonstrate the \name pipeline’s effectiveness in distinguishing high-attentional ERP episodes while mitigating noise artifacts.

\noindent {\bf{\underline{Effectiveness of Verbal Cueing:}}}  
To assess the effectiveness of \emph{cueing}, we analyzed gaze fixations during the navigation task and their distribution across cued and non-cued objects. Figure~\ref{fig:CDFplot} presents the cumulative distribution function (CDF) of the normalized RGB pixel intensity of fixation heatmaps (generated by the GazeRecorder application) on the $x$-axis, where lower intensity values indicate higher visual focus. Our results reveal that over 80\% of cued objects exhibited intensity values below 0.5, whereas only 20\% of non-cued objects reached the same fixation intensity level. Additionally, we conducted a one-way ANOVA with the input method (EEG, EEG Recon, EEG Recon + GSR + PPG) as the categorical independent variable and the proportion of change points corresponding to cued objects as the dependent variable. The analysis revealed significant differences in cued moment detection across input modalities \((F = 10.68, p < 0.0232)\).

\begin{table}
\centering
\caption{Modal Fusion Profiling on Laptop (CPU) based on Time Complexity ($T_{C}$), Noise Proportion ($N_p$) and Accuracy ($D_{Acc}$)}
\scalebox{0.88}{
\begin{tabular}{|l|c|c|c|}
\hline
\textbf{Method} & \textbf{$T_{C}$ (sec)} & \textbf{$N_p$} & \textbf{$D_{Acc}$} \\ \hline
EEG (r) & 0.458 ± 0.016 sec & 0.502 & 0.498 \\ \hline
EEG (re) & 2.38 ± 0.093 sec & 0.396 & 0.604 \\ \hline
EEG (re) + GSR & 3.21 ± 0.121 sec & 0.308 & 0.692 \\ \hline
\textbf{EEG (re) + GSR + PPG }& \textbf{3.86 ± 0.139 sec} & \textbf{0.274} & \textbf{0.726} \\ \hline
\end{tabular}}
\label{tab:modal_fusion}
\end{table}

\noindent {\bf{\underline{Impact of Multi-Modal Fusion:}}}  
Conducted a comparative study to evaluate the effect of multi-modal fusion on detecting high-cognitive ERP episodes. All experiments were performed on a MacBook Pro with an M1 chip, 8GB RAM, and 512GB storage. The results, summarized in Table~\ref{tab:modal_fusion}, show that using raw EEG data yields the lowest detection accuracy (\textbf{0.498}), the fastest processing time (\textbf{0.458} sec), and the highest noise proportion (\textbf{0.502}). Incorporating reconstructed EEG improves accuracy to \textbf{0.604} and reduces noise to \textbf{0.396}, though at the cost of an increased processing time of \textbf{2.38} sec. Further fusion with GSR and PPG enhances detection accuracy to \textbf{0.726} while progressively reducing noise to \textbf{0.274}. However, this improvement increases computational time, reaching \textbf{3.86} sec per iteration. These findings highlight the trade-off between accuracy, noise reduction, and computational efficiency, emphasizing the need to balance these factors for real-time applications.

\begin{table}[t]
\caption{Memory recall of cued and non-cued objects for various environments. DBM - Downtown Baltimore, NYC - New York City.}
\scalebox{0.88}{
\begin{tabular}{|l|c|c|c|c|} \hline
       & Indoor & Campus & DBM  & NYC  \\ \hline
Cued     & 0.98      & 0.86   & 0.64 & 0.39 \\ \hline
Non-Cued & 0.77   & 0.14   & 0.19 & 0.30 \\ \hline
\end{tabular}}
\label{tab:study1-recall}
\end{table}

\noindent {\bf{\underline{Impact of Cueing on Object Recall:}}}  
We evaluated object recall following the navigation task, as summarized in Table~\ref{tab:study1-recall}. Object labels attended to by participants were identified using the state-of-the-art YOLOv4 object detector~\cite{bochkovskiy2020yolov4}. The recall values represent the average normalized recall for both cued and non-cued objects. Key observations include:  
\textbf{(i)} Participants recalled cued objects more accurately (\textbf{66.66\%}) than non-cued objects (\textbf{38\%}), \textbf{(ii)} Although cueing generally improved recall across environments, performance was equal in the most cognitively demanding environment (NYC), which had the highest PAAS score (\textbf{7.6}, see Fig.~\ref{fig:pass_env}), and \textbf{(iii)} The greatest cueing benefit was observed in the least cluttered and most familiar environment—Downtown Baltimore—where recall improved from \textbf{14\%} to \textbf{86\%} (see Table~\ref{tab:study1-recall}), corresponding to a reported PAAS score of \textbf{2.7} across participants.

\noindent \underline{\textbf{Key Takeaways:}} These findings indicate that:  
\textbf{(i)} Consistent with prior studies on working memory, verbal cueing effectively enhances memory recall, and \textbf{(ii)} \name significantly improves the extraction of high-attention moments, demonstrating the effectiveness of its key design features.

\subsection{RQ2: Improving Route Recall via ERPs}\label{sec:s2-recall}  

We assess participants' recall accuracy of their original route as a measure of memory performance. Since the virtual navigation exercise was conducted using third-party web-based tools (Matterport and Google StreetView), precise spatio-temporal trajectories were unavailable. Instead, we evaluate recall accuracy by comparing first-person views captured during the experiments, following prior image-based localization approaches~\cite{indoornav}. These views approximate participants' locations and head orientations, providing a more comprehensive definition of \emph{similarity}. To quantify recall accuracy, we employ appearance-based distance similarity measures under two scenarios:  
\textbf{(a) Case 1}—Pixel intensity similarity between image views at identical time instants across original and re-traced routes, and \textbf{(b) Case 2}—Similarity between image views with a relaxed timeline, allowing variations in speed while maintaining spatial route consistency. For \textbf{Case 1}, we use the Structural Similarity Index (SSIM)~\cite{wang2004image} and Earth Mover's Distance (EMD)~\cite{rubner2000earth} to robustly assess visual similarity and \textbf{Case 2}, we apply the FastDTW algorithm~\cite{salvador2007toward}, a DTW-based method that accounts for temporal misalignment while preserving spatial accuracy.

\begin{figure}[t]
\begin{minipage}{\columnwidth}
\centering
        \begin{subfigure}[t]{0.58\columnwidth}
	   \centering \includegraphics[scale= 0.097]
{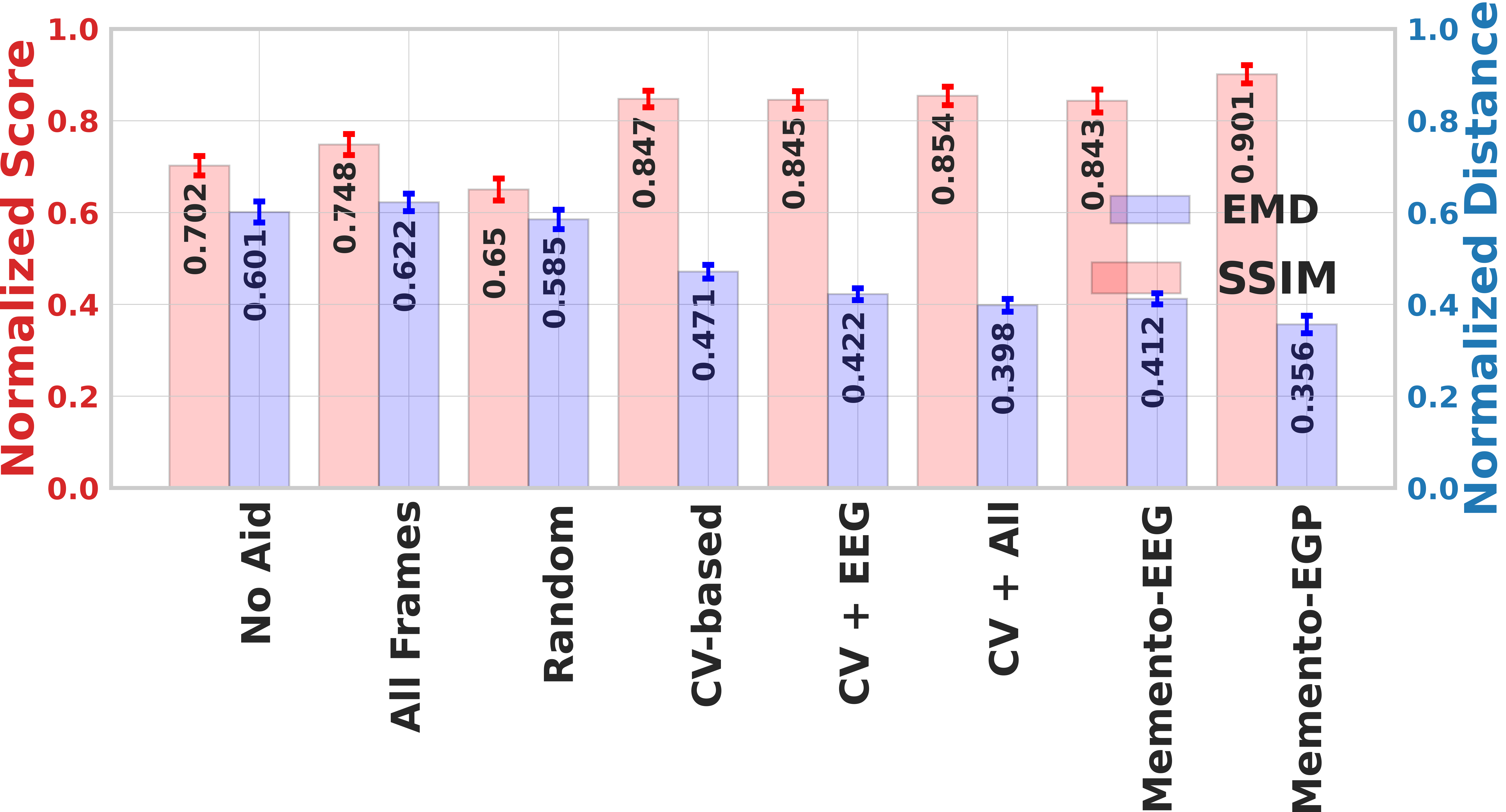}
      \subcaption{\textbf{Case 1}: Both space and time similarity}
        \end{subfigure}%
       \hfill%
        \begin{subfigure}[t]{0.4\columnwidth}
	   \centering \includegraphics[scale= 0.14]{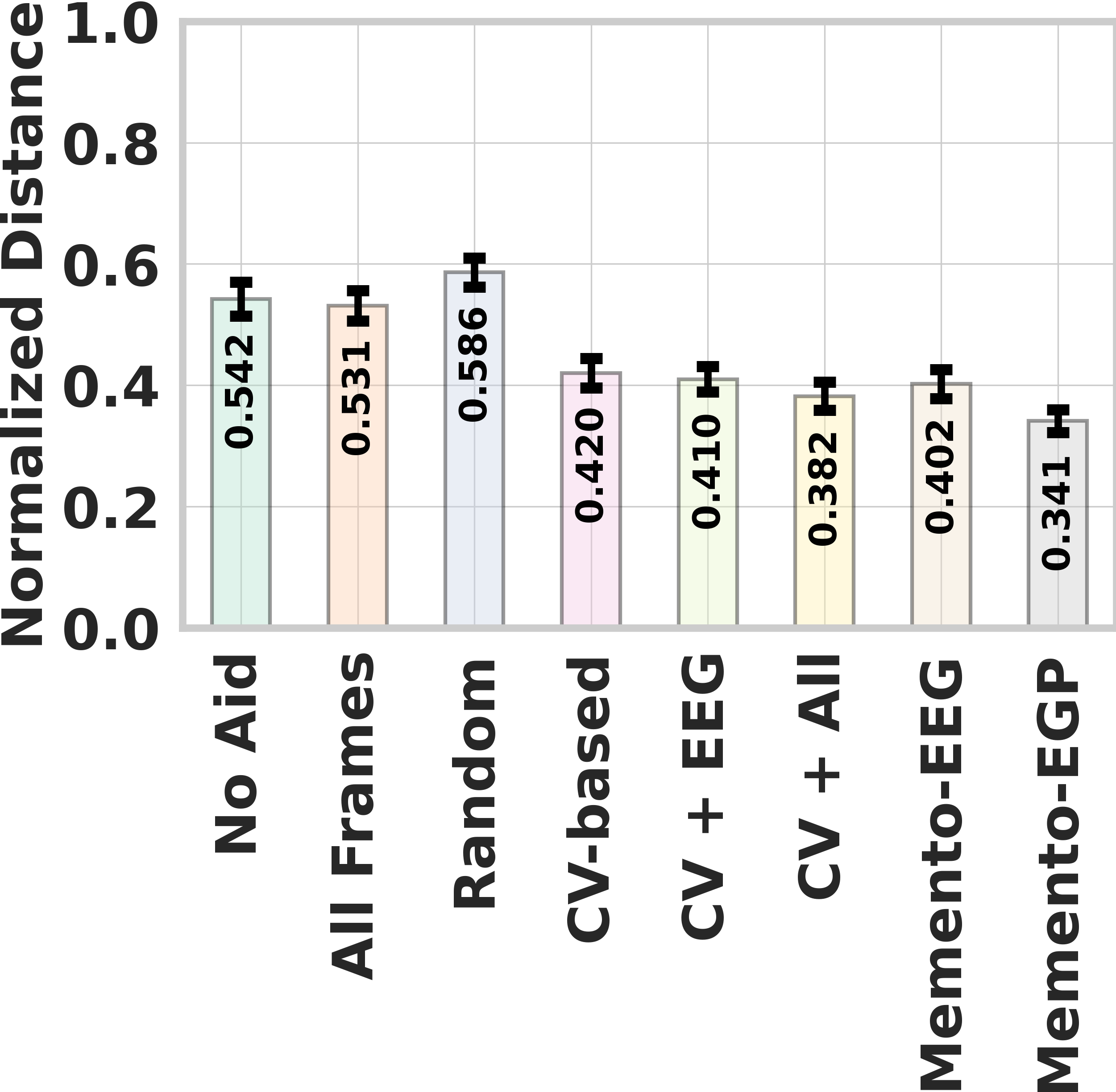}
   \subcaption{\textbf{Case 2}: Spatial similarity with relaxed temporal consistency}
        \end{subfigure}%
    \caption{Retrace accuracy for the eight treatment types {\it (No aid), Memento-EEG, Memento-EGP, CV-based, All frames, CV-EEG, CV-All modalities} and~{\it Random frames} averaged across all subjects and environments for two cases: (a) {\it spatial-based correlation} and (b) {\it temporal-based correlation}, respectively.}
    \label{fig:acc-study2}
\end{minipage}
\end{figure}

The Figure.~\ref{fig:acc-study2} presents the average retrace accuracy across all participants for the three treatment types, leading to the following \emph{key observations:} \textbf{(i)} Fig.~\ref{fig:acc-study2}(a), \names-EEG achieves performance comparable to the CV-based approach (SSIM: 0.843 vs. 0.847). However, \names-EGP, which integrates all physiological signals, demonstrates the highest effectiveness, yielding an approximate \textbf{6\%} improvement, \textbf{(ii)} Selective cueing significantly enhances recall over free recall or random selections, resulting in an improvement of \textbf{15.3\%--25.1\%}, with consistent results observed using the EMD measure, \textbf{(iii)} The improvement in FastDTW is less pronounced due to its relaxed definition of retrace accuracy (Fig.~\ref{fig:acc-study2}(b)). For instance, \names-EGP improves recall by \textbf{20.01\%} compared to no memory aid, as reflected in a reduced normalized distance. However, physiological sensing-based cueing remains as effective as the CV-based approach \textit{(CV: \textbf{0.420} vs. \names-EEG: \textbf{0.402})}, and \textbf{(iv)} Further filtering of CV-based selections using physiological signals enhances recall; however, the improvements are not significant enough to warrant the sensing overheads of all modalities.

\noindent {\bf{\underline{Effectiveness of Cueing:}}} \label{sec:s2-cueing}  
While retrace accuracy measures the similarity between the original and retraced routes, it does not directly evaluate the effectiveness of the \emph{cueing} process. We assess the similarity between \emph{visual cues} (i.e., sub-selected frames) and the retraced routes to address this. Table~\ref{tab:cueing-effec} presents the mean and standard deviation of the \textbf{SSIM} and Pearson correlation coefficient~(\textbf{PCC}) scores across different treatment types. Consistent with previous findings (Fig.~\ref{fig:acc-study2}), \name-EGP achieves the highest similarity scores (\textbf{0.85} and \textbf{0.73}), whereas \name-EEG and the CV-based approach yield comparable results (\textit{[0.67, 0.59]} vs. \textit{[0.70, 0.63]}, respectively). Lower similarity scores for \textit{Random} and \textit{All Frames} treatments (\textit{[0.40, 0.42]} vs. \textit{[0.54, 0.48]}) indicate a significant divergence between the provided cues and participants' actual retraced routes, highlighting the ineffectiveness of these treatments.

\begin{table*}
\caption{Similarity scores between the \emph{visual cues} and \emph{retraced routes}.}
\scalebox{0.98}{
\begin{tabular}{|l|c|c|c|c|c|c|c|}
\hline {\bf Metrics} & Memento-EGP & CV-All & CV-EEG  & Memento-EEG & CV & All Frame & Random   \\ \hline
{\bf SSIM}    & 0.85 ($\pm$ 0.0345)     & 0.78 ($\pm$0.0313)   &  0.72 ($\pm$ 0.0298) & 0.70 ($\pm$ 0.0312) & 0.67 ($\pm$ 0.0281) & 0.54 ($\pm$ 0.0291) & 0.40 ($\pm$ 0.0229)\\ \hline
 {\bf PCC} & 0.73 ($\pm$ 0.0218)     & 0.65 ($\pm$0.0199)   &  0.58 ($\pm$ 0.0208) & 0.63 ($\pm$ 0.0211) & 0.59 ($\pm$ 0.0239) & 0.48 ($\pm$ 0.0213) & 0.42 ($\pm$ 0.269)\\ \hline
\end{tabular}}
\label{tab:cueing-effec}
\end{table*}

\begin{figure}
\begin{minipage}{\columnwidth}
\centering
        \begin{subfigure}[t]{0.33\columnwidth}
	   \centering \includegraphics[scale= 0.30]{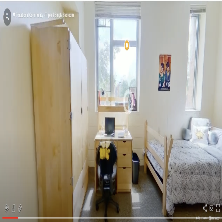}
      \subcaption{Ground truth frame}
        \end{subfigure}%
       \hfill%
        \begin{subfigure}[t]{0.35\columnwidth}
	   \centering \includegraphics[scale= 0.30]{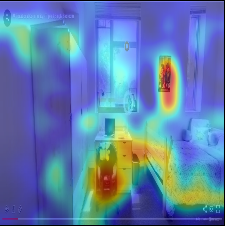}
   \subcaption{CV-based~(AmNet) attention distribution}
        \end{subfigure}%
        \hfill%
        \begin{subfigure}[t]{0.28\columnwidth}
	   \centering \includegraphics[width=2.33cm, height=2.33cm]{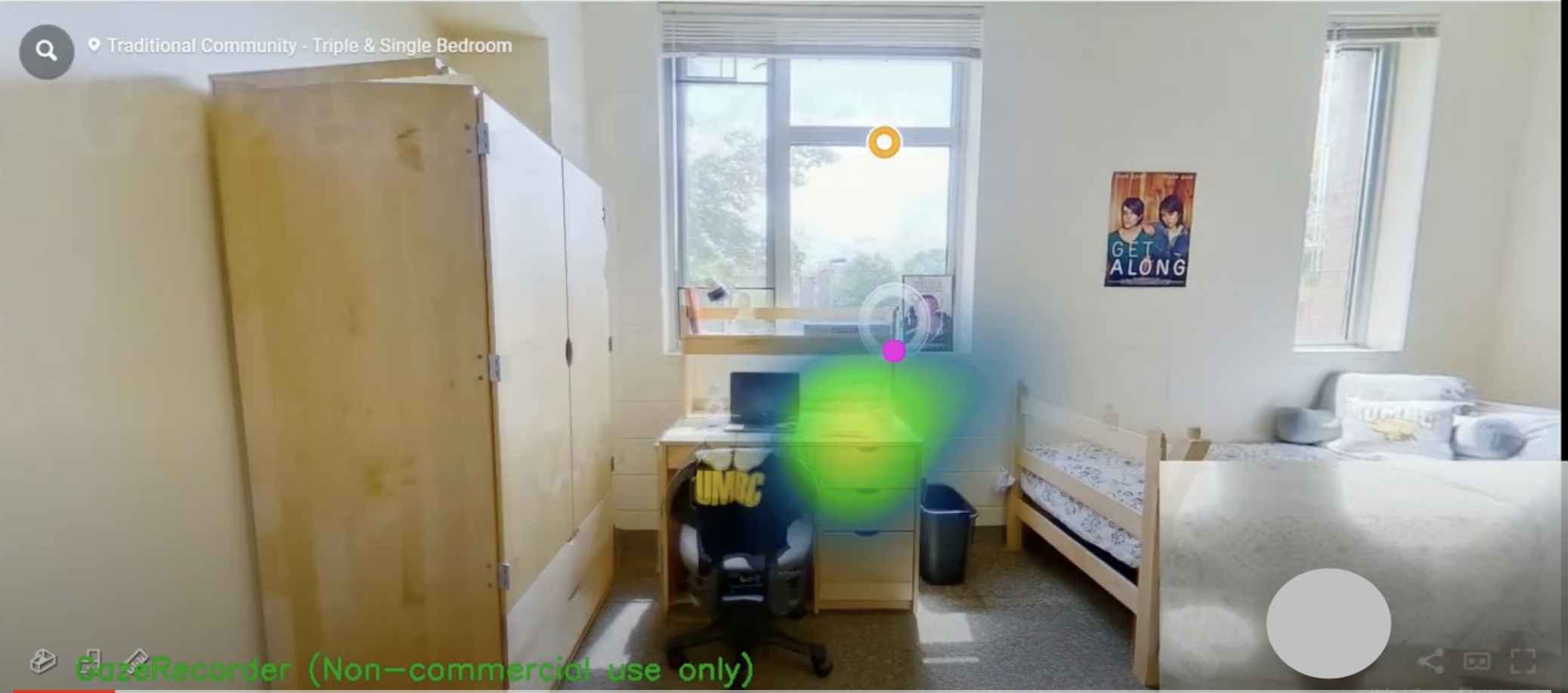}
   \subcaption{Gaze-based attention distribution}
        \end{subfigure}%
    \caption{{Qualitative comparison insights between the human visual salient attention frame distribution and the CV-based salient attention frame distribution.}}
    \label{fig:cv-study2_comparison_fixation}
\end{minipage}
\end{figure}

\noindent {\bf{\underline{Assessment of Reviewing Load.}}} \label{sec:s2-load}  
While our results indicate that the physiological sensing-based approach performs as well as, or better than, the computer vision-based approach in the route recall task, we further assess the \emph{efficiency} of the cueing process by analyzing the reviewing load. Given the strict 1-minute review limit, we compare the number of frames selected for cueing by both the CV-based and \name-EGP approaches. At a 67.7\% memorability threshold, the CV-based approach selects \textbf{71--76\%} of the total frames from the original explorations, amounting to approximately \textbf{435--478} frames per episode. In contrast, the \name framework, leveraging pre-processed multimodal sensors (\textbf{Memento-EGP}) and EEG sensors only (\textbf{Memento-EEG}), selects significantly fewer frames—approximately \textbf{228--257} and \textbf{295--324} frames, respectively. This reduction corresponds to a \textbf{16--46\%} decrease compared to the CV-based approach. As discussed in Section~\ref{sec:s2-stats}, these findings align with participant ratings in the post-study questionnaire.

Furthermore, employing the CV-based approach (AmNet) for memory recall reveals a notable disparity in the retrieval of relevant information, as shown in Fig.~\ref{fig:cv-study2_comparison_fixation} (b) and (c). The comparison indicates that humans recall fine-grained, specific details, whereas the CV-based AmNet system, despite its efficiency, integrates both relevant and irrelevant information, potentially leading to cognitive overload misaligned with human memory priorities. Our model \textit{\textbf{personalizes}} memory recall by prioritizing salient features and cue objects tailored to individual user preferences. In contrast, the CV-based AmNet system follows a generalized approach, presenting broad, non-personalized information that lacks human-centric relevance. This fundamental difference aligns \namei~with users’ selective attention and memory patterns—focusing on meaningful events rather than overwhelming them with non-personalized, generic semantic cues, as shown in Fig.~\ref{fig:cv-study2_comparison_fixation} (b). As a result, AmNet tends to \textbf{increase cognitive load}, i.e., perceived load during memory retrieval, whereas \namei~effectively reduces review effort and cognitive load by capturing and leveraging user-specific cues.

\subsubsection{Statistical Analysis:}
\label{sec:s2-stats}

We analyzed the impact of various factors on route recall using one-way ANOVA analyses.

\noindent{\textbf{\underline{Impact of Environment Complexity:}}}  
We conducted a one-way ANOVA with retrace accuracy as the dependent variable and environment complexity as the independent variable (Campus - low, Indoor - medium, City - high). More complex environments contain a higher variety and frequency of objects. The results showed statistically significant differences ($F=13.763$, $p=0.0282$), consistent with our observations from Study 1. \name achieved the lowest performance improvement in the most complex environment (City). The average PAAS scores for each environment were: Indoor: $4.13 \pm 0.22$, Campus: $3.25 \pm 0.26$, and City: $6.34 \pm 0.28$. 

\begin{figure}[t]
\begin{minipage}{\columnwidth}
\centering
        \begin{subfigure}[t]{0.48\columnwidth}
	   \centering \includegraphics[scale= 0.17] {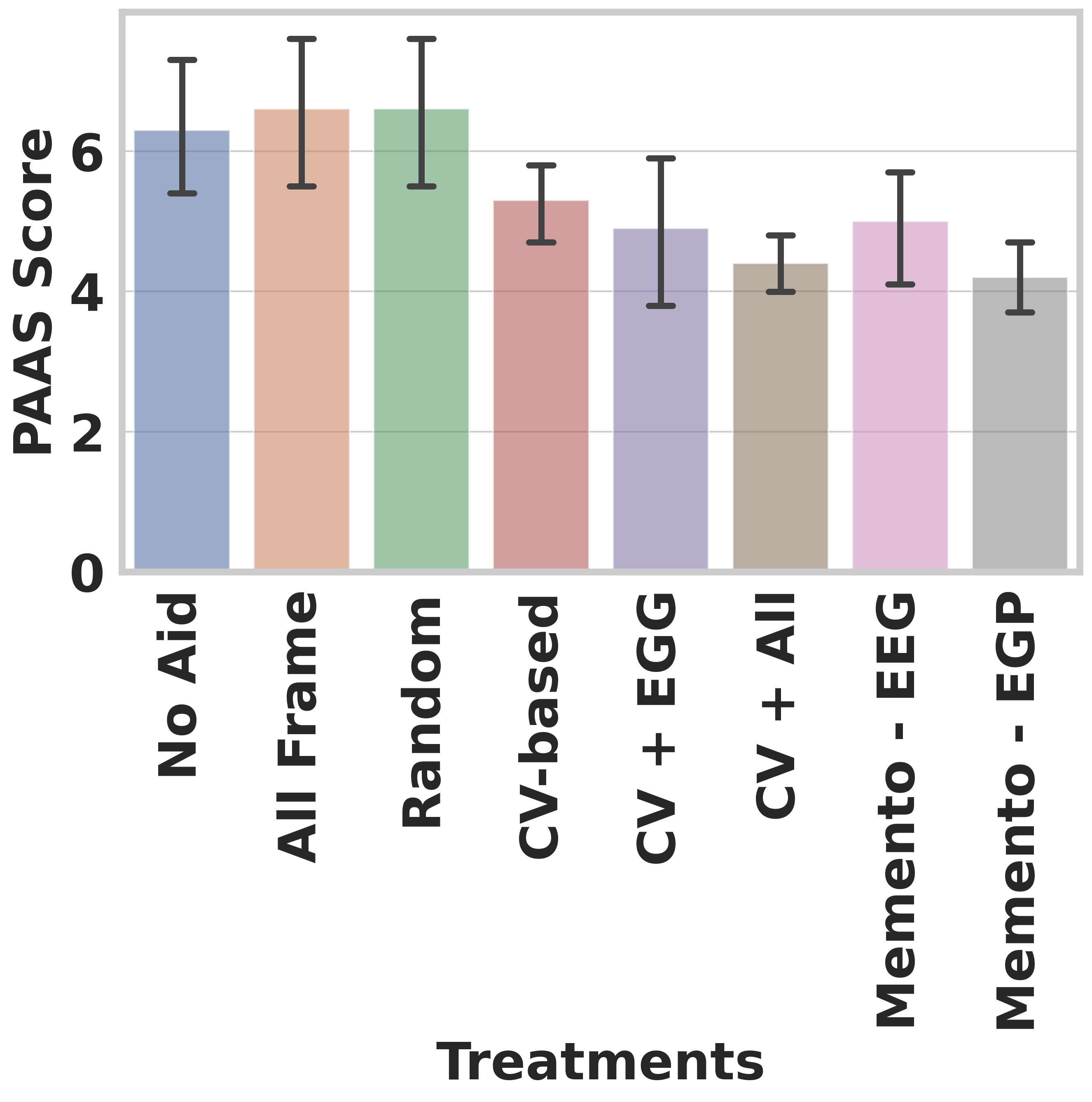}
      \caption{PAAS Score} 
      \label{fig:PAAS Score}
        \end{subfigure}%
        \hfill%
        \begin{subfigure}[t]{0.45\columnwidth}
	   \centering \includegraphics[scale= 0.17]{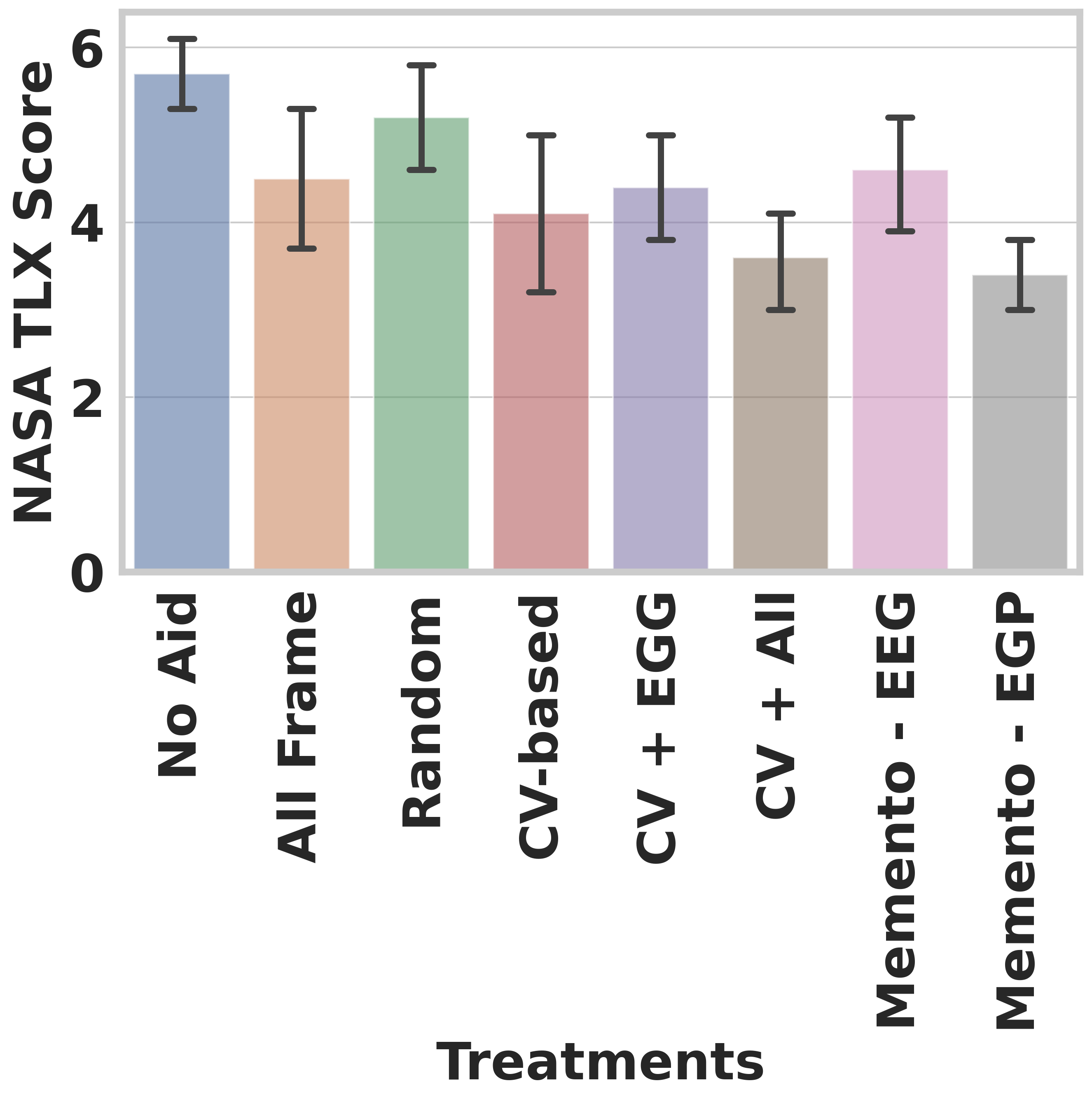}
    \caption{NASA TLX Score}
\label{fig:NASA Score}
        \end{subfigure} 
\caption{Dependent variables with respect to the seven types of memory treatments {\it (No aid, Reconstructed EEG-based selected frames, All frames, AmNet-based selected frame, AmNet-based selected frames and EEG, AmNet-based selected frames and Reconstructed EEG and other modalities and Random frames.}}
\end{minipage}
\end{figure}

\noindent{\textbf{\underline{Perceived Load of the Navigation Task:}}}  
A one-way ANOVA was conducted to evaluate participants' perception of the overall cognitive load with treatment type as the independent variable and the PAAS score as the dependent variable. Statistically significant differences were observed ($F=10.902$, $p=0.0301$), as shown in Fig.~\ref{fig:PAAS Score}. Participants reported the lowest cognitive load with \textbf{Memento-EGP} (score $\approx 3.9$), outperforming all baselines. To further assess the cognitive load associated with the highlighter tool, another one-way ANOVA was performed with treatment type as the independent variable and the NASA TLX score as the dependent variable, revealing statistically significant differences ($F=10.843$, $p=0.0179$) (see Fig.~\ref{fig:NASA Score}). Participants rated \textbf{Memento-EGP} as the least cognitively demanding. Moreover, \name received the highest confidence rating ($4.10 \pm 0.18$) out of 5, surpassing the no-memory-aid baseline ($2.34 \pm 0.24$) and the CV-based baseline ($3.57 \pm 0.17$) and these findings are consistent as discussed in section~\ref{sec:s2-load}.

\noindent{\textbf{\underline{Comparative Computational Analysis of \name:}}}~\label{sec:discussion} While our primary focus has been on establishing the feasibility of using COTS physiological sensors for memory augmentation, the choice of sensing modalities significantly impacts system performance. To evaluate computational efficiency, we compared the runtime of our \textit{Memento} framework against the CV-based AMNet approach. Experiments were conducted on a MacBook Pro (M1 chip, 8GB RAM, 512GB storage), demonstrating that our proposed techniques can efficiently run on typical mobile and edge platforms. Results indicate that our CPD-based \textit{Memento} framework, applied to simple time-series data, requires significantly less processing time per user stimulus session compared to AMNet, which processes richer first-person visual data through deep neural networks. \textit{Memento} achieves an average execution time of \textbf{3.86 ± 0.09 sec}, whereas the AMNet (CV-based) approach requires \textbf{15.35 ± 0.16 sec}. This represents an approximate \textbf{75\%} reduction in processing time, underscoring the computational efficiency of physiological sensing-based cueing for real-time inference.

\noindent{\textbf{\underline{Key Takeaways:}}}
The findings from Study 2 demonstrate that visual cueing significantly enhances navigation task recall. While EEG-based and CV-based approaches showed comparable performance, integrating additional physiological modalities (GSR and PPG) further improved cueing effectiveness. Notably, participants rated the physiological sensing-based approach more favorably, indicating a lower perceived cognitive load, as it enabled the capture of more \textbf{personalized cues} compared to other methods.

\section{Conclusion}
In this work, we proposed and evaluated a first-of-its-kind system which extracts moments or highlights from a user's short-term episodes to cue those important moments during visuospatial working memory tasks, intending to improve the memory recall of those moments. Through a system implementation and two separate user-studies, we establish that improving one's recall with intelligent cueing is possible during visual search and wayfinding navigation. We show that participants receiving visual cues from \namei~improved route recall by \textbf{20–23\%}, reduced cognitive load and review time by \textbf{46\%}, and achieved a \textbf{75\%} faster computation time~(\textit{3.86 secs} vs. \textit{15.35 secs})~compared to vision-based cue selection.

\section{Limitation and Future Directions}

In this work, we primarily employed time- and frequency-domain features, however, one limitation of our current approach is the lack of integration of non-linear physiological indices. As part of future work, we plan to incorporate additional non-linear measures\cite{laborde2017heart} to enable more comprehensive physiological interpretations. Furthermore, \textit{Memento's} strong empirical results align with retrieval practice~\cite{roediger2006test}, which emphasizes personalized memory augmentation through active recall, and dual-coding~\cite{paivio1991dual}, which underscores the benefits of combining visual and verbal cues. In future work, we aim to enhance human-computer interaction through immersive environments~\cite{han2021mobile, saeghe2022augmenting, hallgarten2024gears}, extending beyond navigation to applications in memory recall student learning~\cite{laporte2023laureate} and older adults~\cite{lindgren2024adapt}.

\section{Acknowledgement} \label{sec:ack}
This work has been partially supported by NSF CNS EAGER Grant $\#2233879$, U.S. Army Grant~$\#W911NF2120076$, U.S. Army Grant \#W911NF2410367, ONR Grant~$\#N00014$-$23$-$1$-$2119$, NSF CAREER Award~$\#1750936$ and NSF REU Site Grant $\#2050999$. In addition, the authors would like to thank all our volunteers who provided the data at the University of Maryland Baltimore County.

\bibliographystyle{ACM-Reference-Format}
\bibliography{Reference}
\end{document}